# Plasmon-interband hybridization and anomalous production of hot electrons in aluminum nanoantennas

*Jérôme Martin,*[*,†] *Oscar Avalos-Ovando,*[¶] *Thomas Simon,*[†] *Gabriel Arditi,*[‡] *Florian Lamaze,*[†] *Julien Proust,*[†] *Luiz Tizei,*[‡] *Zhiming Wang,*[1,2] *Mathieu Kociak,*[‡] *Alexander O. Govorov,*[¶] *Odile Stéphan,*[‡] *and Davy Gérard*[*,†]

†Lumière, nanomatériaux, nanotechnologies (L2n), UMR CNRS 7076, Université de Technologie de Troyes, Troyes 10004, France

‡Laboratoire de Physique des Solides, Bâtiment 510, UMR CNRS 8502, Université Paris Saclay, Orsay 91400, France

¶Department of Physics and Astronomy and Nanoscale and Quantum Phenomena Institute, Ohio University, Athens, Ohio 45701, United States

1 Institute of Fundamental and Frontier Sciences, University of Electronic Science and Technology of China, Chengdu 61173, China

2 Shimmer Center, Tianfu Jiangxi Laboratory, Chengdu 641419, China

ABSTRACT

Strong coupling typically occurs between two separate objects or between an object and its environment (such as an atom and a cavity). However, it can also occur between two different excitations within the same object, a situation that has been much less studied. In this study, we observe strong coupling between localized surface plasmon resonances and the interband transition in aluminum nanorods, as evidenced by optical spectroscopy and electron energy loss spectroscopy, and corroborated with numerical simulations. Strong coupling is observed between the interband transition and multiple orders of the surface plasmon mode, including dark ones. We also obtain experimental maps of the hybrid modes at the nanoscale. In each case, the associated Rabi energy, which corresponds to the energy splitting between the two polaritonic branches, is obtained. Moreover, a dedicated numerical model was employed to calculate the hot electron generation rate in the nanorods. The calculations demonstrate that efficient generation of hot electrons can be achieved in the near-infrared region, when the interband transition is strongly coupled with a plasmon resonance. This high generation rate stems from the hybrid nature of the mode, as its plasmonic component provides a high absorption cross-section, while the interband transition part ensures efficient conversion to hot electrons. Consequently, aluminum nanorods represent an efficient source of hot electrons in the visible and near-infrared regions, with potential applications in local photochemistry, photodetection, and solar energy harvesting.

Strong coupling is defined as a situation in which two coupled oscillators are no longer adequately described by the eigenstates associated with the initial (uncoupled) oscillators. The paradigmatic situation for strong coupling is an atomic transition inside a high-quality optical cavity, which can be rigorously described by cavity quantum electrodynamics (cQED).[1] In this case, the coupled system becomes a hybrid system, described by light-matter states known as polaritons.[2] This hybridization leads to a change in the eigenstates and energies of the system. In the weak coupling regime, due to the low energy exchange rate between the oscillators, each of them can be described by its own eigenstates. However, as the coupling strength increases, the new energy levels can diverge significantly from the original ones and new eigenstates emerge, which are a superposition of the uncoupled states. Although strong coupling is rooted in the quantum world, it is noteworthy that a classical picture with mechanical oscillators can still capture many of the characteristic features of strongly coupled systems, such as energy splitting and avoided crossing.[3]

Following the pioneering cQED experiments of the 1980s, the potential for strong coupling in micro- and nanophotonic systems, such as optical microcavities, was identified through the advancement of fabrication technologies. This led to the observation of strong coupling at room temperature.[4] Of particular interest are emitters coupled to nanostructures that sustain surface plasmon resonances.[5] Examples include noble metal nanoparticles coupled to J-aggregates,[6–8] dye molecules,[9] quantum dots,[10] photochromic molecules,[11] ZnO excitons,[12] 2D materials[13–17] and single molecules in ultra-small plasmonic cavities.[18] In comparison to cQED, nanophotonic systems offer a simpler platform for studying strong coupling. In addition to operating at room temperature, nanoparticles sustaining localized surface plasmon resonances (LSPRs) act as open cavities, facilitating the coupling with emitters. Consequently, nanophotonics represents an exceptionally promising platform for the observation and study of strong coupling.[19]

In 2011, Pakizeh[20] investigated theoretically the plasmonic properties of metals exhibiting a spectrally localized interband transition (IT), employing a dielectric function based on the Drude-Lorentz model. He demonstrated that the spectral positions of the LSPR and the IT exhibited an avoided crossing behavior, which is indicative of a hybridization of the two modes. Subsequently, several experimental confirmations of this prediction have now been published, in aluminum[21,22] and nickel[23–25] nanostructures. What makes this phenomenon particularly intriguing is that the Lorentzian oscillator that couples with the LSPR is not an external electric dipole, but rather the nanoparticle itself via its dielectric function. In other words, *the particle couples to itself*, thereby creating hybridized states between two different excitations in the same object, as illustrated in **Figure 1a**. Strong coupling between two different excitations of the same nature (two different surface plasmons) has also been observed.[26] It is noteworthy that so far, plasmon-IT hybridization was only observed in materials with a spectrally localized IT, akin to a transition dipole, such as aluminum and nickel. This phenomenon has not been observed in gold, for instance.[21] It should be noted that it has recently been proposed that similar self-coupling behavior could also be observed in high refractive index dielectrics exhibiting Mie resonances.[27]

From a microscopic perspective, a LSPR excitation is, in essence, a coherent superposition of low-energy electrons near the Fermi level.[28] After a few femtoseconds, this excitation decays either via a radiative process (photon scattering, which is the dominant mechanism for larger nanoparticles), acoustic phonon emission, or via the creation of short-lived, high-energy electron-hole pairs.[29,30] These energetic carriers form a non-thermal distribution for a few hundreds of femtoseconds before thermalization into a Fermi-Dirac distribution exhibiting a characteristic electronic temperature and, ultimately, increase the lattice temperature.[31,32] These non-thermal, energetic electrons are known as hot electrons (HEs). HE generation has become a hot topic[33] owing to its potential applications notably in photocatalysis,[34,35] solar energy harvesting,[36,37] photodetection,[38,39] and thermionic emission.[40] The generation of HEs from the decay of the collective excitations involves two main channels: interband and intraband.[41] These channels are sketched in **Figure 1b** as colored arrows. A metal nanoparticle illuminated at its IT frequency efficiently generates interband HEs,[42] sketched as red arrows in **Figure 1b**. The efficiency of generation of intraband HEs in metals is typically lower since one needs to break the momentum conservation in the optical electronic transition by using surfaces (Kreibig's mechanism[43], green arrows in **Figure 1b**), or phonons[44-46] (orange arrow in Figure 1b). In previous works[47,48] an unusually high efficiency of HE generation in silver-coated gold nanorods was observed, which is related to the IT in silver. However, the IT transition in silver is in the UV, which hinders many potential applications for solar light-to-HE conversion, which are related mostly to visible light. Aluminum is particularly promising in this respect since it possesses its IT transition and the corresponding HE production channel in the visible/near-infrared range.

In this article, we present a numerical and experimental study of the coupling between IT and plasmon resonances sustained by aluminum nanorods, as well as a theoretical exploration of their ability to generate HEs in the visible range. In contrast with previous experimental works on self-strong coupling, tuning the nanorods length and hence their LSPR energies permits to systematically explore the coupling between the IT and multipolar (both bright and dark) plasmon resonances. Moreover, we employ Electron Energy-Loss Spectroscopy (EELS) to map the resulting hybrid modes at the nanoscale. We then demonstrate, through theoretical analysis, that the strong coupling between surface plasmon and IT enhances the generation of hot electrons. This enhancement is attributed to the hybrid nature of the coupled mode, where the plasmonic component contributes with its large absorption cross-section, while the IT component facilitates efficient hot electron generation.

In the following, we focus on aluminum nanorods as they are an optimal choice of geometry for several reasons. First, aluminum exhibits a localized IT around 1.5 eV with a linewidth of approximately 200 meV.[49] As shown by the red arrows in **Figure 1b**, this spectrally localized absorption band originates from transitions occurring between near-parallel bands in the vicinity of the K and W symmetry points.[50,51] Second, aluminum can sustain LSPR over a broad spectral range, from the UV to the infrared.[52] Third, nanorods sustain multiple-order surface plasmon resonances, whose spectral position can be easily tuned by changing the length of the nanorod[53,54] - similar to a classical optical cavity. In the next section, we initiate our examination with the presentation of numerical simulations. These simulations provide an overview of the studied phenomenon.

## RESULTS AND DISCUSSION

### Numerical calculations

We consider Al nanorods with lengths *L* ranging from 100 to 900 nm, where width and height are held constant at 40 nm. The substrate consists of a thin (15 nm) silicon nitride ($Si_3N_4$) membrane. A sketch of the structure is provided in **Figure 1a**. Finite-difference time-domain (FDTD) calculations were performed on this structure, using Ref. [55] to model the complex refractive index of aluminum (see the Methods section for details of the calculations).

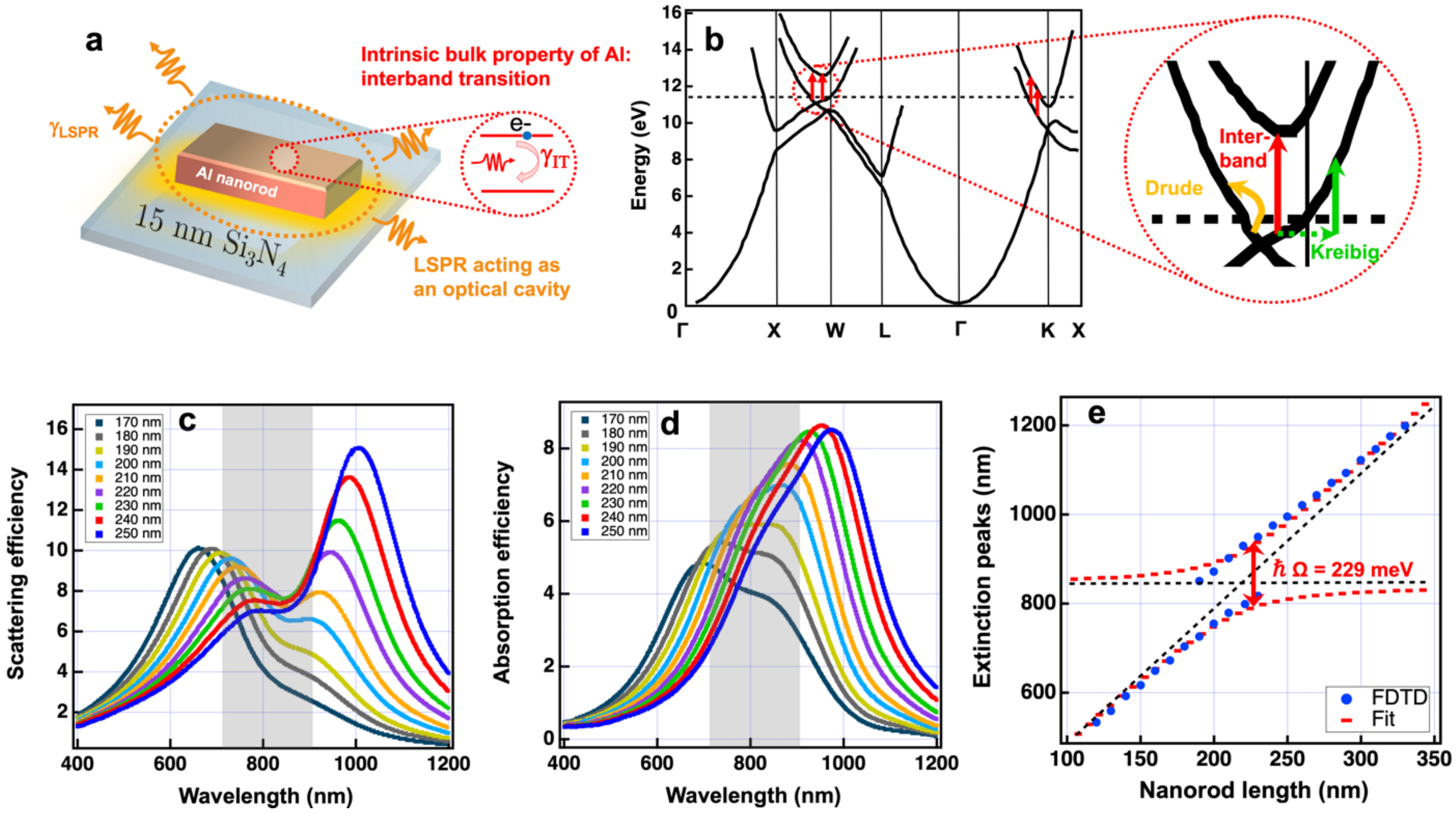

**Figure 1. a** Diagrammatic representation of the hybridization between the interband transition and the dipole LSPR in an aluminum nanoantenna. **b** Schematics of the band structure of Al. The dotted horizontal line corresponds to the Fermi level and the red vertical arrows indicate possible interband transitions. The left panel is a zoom-in around the W symmetry point near the Fermi level. The colored arrows illustrate different possible transitions: interband transition (red arrow), Drude intraband transition (orange arrow), and surface-assisted intraband transition (Kreibig's mechanism, green arrows). **c** FDTD-calculated scattering efficiency spectrum of an aluminum nanorod as a function of its length. The area within the delineated region in gray corresponds to the IT region. **d** Same, for the absorption efficiency. **e** Evolution of the wavelength of the extinction peaks maxima with length. The red dotted lines correspond to a numerical fit of the data using Eq. 1, and the black dotted lines are the input values of $\omega_{SP}$ and $\omega_{IT}$ entered in Eq. 1.

**Figure 1** depicts the FDTD-calculated scattering (Figure 1c) and absorption (Figure 1d) efficiencies as a function of wavelength for different nanorod lengths. We first focus on a limited range of nanorod lengths (100 to 340 nm) corresponding to the dipole LSPR. Two distinct peaks are observed, situated on either side of the IT peak wavelength, at longer and shorter wavelengths, respectively. The presence of two peaks can be attributed to two polaritonic modes, resulting from the coupling between the dipole mode excited along the long axis of the nanorod, and the IT. This phenomenon can be modeled by using a Drude-Lorentz function for the permittivity of aluminum.[22] The IT is associated with a Lorentzian oscillator, that couples to the dipole plasmonic mode supported by the nanoantenna itself. This phenomenon is one of the primary characteristics of systems exhibiting strong coupling, wherein the emergence of two new eigenmodes is observed. This effect is particularly pronounced at zero wavelength detuning between the two initial resonances. Furthermore, the scattering and absorption spectra exhibit different behaviors. For rod lengths of approximately 200–230 nm, where the coupling with the interband is the strongest, a local minimum (dip) in scattering efficiency is observed at the IT wavelength. This is indicative of an avoided crossing between two resonances, as it will be demonstrated in the following section. The avoided crossing behavior is less pronounced in the absorption spectra, where instead of a clear dip a shoulder in the absorption peak is observed. The observation of a more pronounced dip in the scattering spectrum can be attributed to the size of the nanorods. In general, the extinction efficiency of plasmonic nanoparticles is dominated by absorption for small sizes and by scattering for larger particles.[56] For rods within the length range studied here, scattering overcomes absorption. We emphasize that in purely scattering spectra, true Rabi splitting can sometimes be confused with far-field interference effects.[19] In the following we will exclusively present extinction spectra, which encompass both absorption and scattering.

Strong coupling is also evidenced in **Figure 1e**, which shows the peak maxima extracted from the FDTD calculations plotted as a function of the nanorod length. This dispersion curve shows two branches which can be associated with the upper branch (UB) and lower branch (LB) self-coupled polaritons. The splitting in the branches is attributed to the strong coupling between the IT and the dipole plasmonic mode. When the LSPR and the IT are largely detuned, the branches become indistinguishable from uncoupled LSPR modes. These numerical data are then compared with a classical coupled oscillator model:[23,57]

$$\omega_{\pm} = \frac{1}{2}\left(\omega_{IT} + \omega_{SP} \pm \sqrt{\Omega_R^2 + (\omega_{IT} - \omega_{SP})^2}\right) \qquad (1)$$

where $\omega_{IT}$ and $\omega_{SP}$ are the initial (before coupling) angular frequencies of the IT and the LSPR, respectively and $\Omega_R$ is the Rabi angular frequency. We emphasize that the model underlying Eq. 1 does not incorporate loss. While the system under observation does manifest losses, a point that will be discussed in more detail later, Eq. 1 is sufficient for the purpose of fitting the two observed branches[5] (see also section 7.8 in Ref. [58]) and to retrieve the Rabi angular frequency $\Omega_R$. In order to obtain a dispersion relation $\omega_{SP} = f(L)$ in the absence of interband transitions (where $L$ is the nanorod length), we assumed that the wavevectors of the plasmonic resonances are given by $k_{SP} = m\frac{\pi}{L}$, with $m$ an integer corresponding to the order of the mode,[54] and we fitted the dispersion relation with a polynomial. More details about this procedure, including fitting parameters for the dispersion relation, are given in the Supplementary Information. The results from this procedure appear as the red dashed lines in **Figure 1d**. It is also possible to extract the splitting energy $\hbar\Omega_R$ directly from the fitting procedure. In the case illustrated in **Figure 1d**, we obtain the value of $\hbar\Omega_R = 229$ meV for the splitting energy.

We have also numerically studied the coupling between IT and higher order plasmonic modes (see Supplementary Information, **Figure S1**). In particular, we carried out FDTD simulations for nanorods of longer lengths, ranging from 700 nm to 1000 nm, sustaining an octupole mode, which is the first high-order bright mode (the quadrupole mode cannot be coupled to the radiative continuum, due to its zero net dipole moment). These calculations showed that the plasmonic octupole mode and IT are also coupled with a lower Rabi splitting compared to the dipole mode, $\hbar\Omega_R = 153$ meV. Interestingly, the coupling strength seems to decrease for higher order modes. This effect will be confirmed and discussed in the following experimental part.

We also performed numerical simulations of the extinction cross-section using the COMSOL software. The model, including the values for the permittivity of Al, is presented in the Supplementary Information in **Figure S2**. We obtained results analogous to those from Lumerical (**Figure S3**). The COMSOL modeling employed in this study facilitates the examination of the structure of the extinction cross-section. The final section of this article will present a decomposition that divides the total cross-section into its scattering and absorption components, and further subdivides the absorption into Drude, interband, and surface-scattering mechanisms. The dissipation channel attributed to surface-scattering (intraband hot-electrons) is found to be negligible in this context, owing to the relatively large size of the Al nanorods.[45] However, we will demonstrate below that the interband hot-electron processes in our Al-based systems are highly active.

**Optical spectroscopy**

Extinction spectroscopy was performed on arrays of aluminum nanoantennas. The arrays were fabricated by electron beam lithography with the geometries described in the preceding section (see Methods for details of sample fabrication). It should be noted that non-regularly spaced nanorods (see the inset of **Figure 2c**) were employed to prevent any diffraction-related effect from influencing the resulting spectra. **Figure 2** depicts the results of extinction measurements conducted on different arrays, wherein the length of the nanorods ranges from 100 nm to 350 nm. The nanorods then exhibit a plasmonic dipole mode that spans the spectral region between 400 and 1200 nm. The hybridization of LSPR and IT is evidenced by the splitting of the plasmonic branches. Subsequently, the experimental data are compared with the calculated data using Equation 1, as indicated by the red dotted lines in **Figure 2**. The agreement between the data and the model is very good. The resulting Rabi splitting is $\hbar\Omega_R =$ 225 meV. This value is in close agreement with the previously obtained numerical calculation result of $\hbar\Omega_R =$ 229 meV. We also note that contrary to what is observed in the numerical calculations, there is no difference in the peak intensities between the shortest and longest rods. This is a compensating effect resulting from the higher density of nanoantennas within the arrays of shorter rods in comparison to arrays of longer rods. Extinction measurements conducted on longer nanorods (approximately 600 nm in length) revealed that the octupole mode also couples with the IT. The data obtained are similar to those shown for the dipole mode in **Figure 2** (see the discussion below), but with a Rabi energy splitting $\hbar\Omega_R =$ 164 meV, which is significantly lower than in the case of the dipole mode. In contrast, the quadrupole mode (dark mode) could not be observed experimentally using extinction spectroscopy.

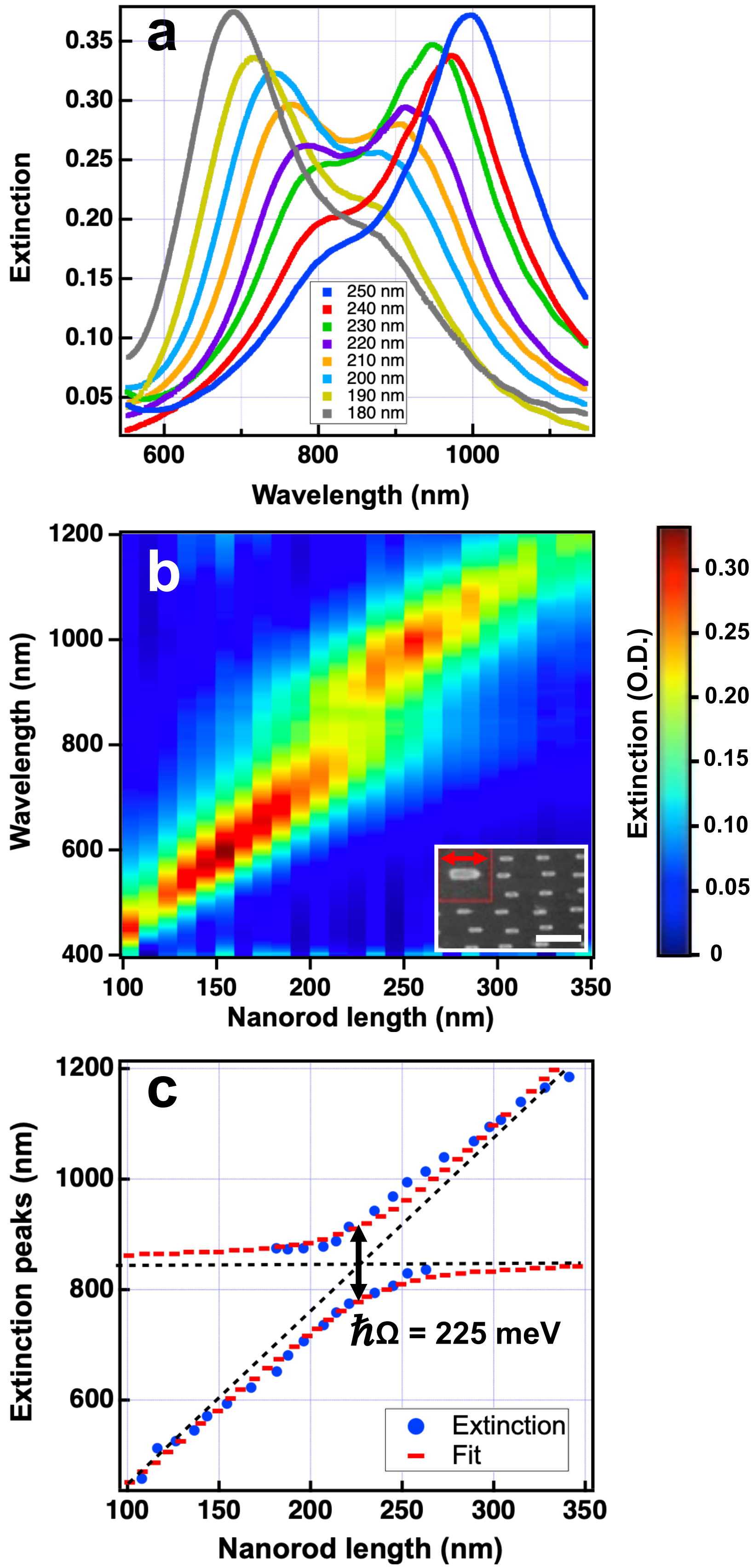
a
Extinction
Wavelength (nm)
250 nm
240 nm
230 nm
220 nm
210 nm
200 nm
190 nm
180 nm
b
Wavelength (nm)
Nanorod length (nm)
Extinction (O.D.)
c
Extinction peaks (nm)
Nanorod length (nm)
ħΩ = 225 meV
Extinction
Fit

**Figure 2.** Optical extinction spectroscopy. **a** Experimental extinction spectra measured on Al nanorods arrays with varying lengths. **b** Extinction efficiency maps measured on various Al nanorods arrays as a function of the length of the rods and the wavelength. Inset: SEM image of a representative nanorod array, scale bar 500 nm. The arrow represents the polarization direction. **c** Evolution of the position of the extinction peaks maxima with length (blue dots). The red dotted lines correspond to a numerical adjustment of the data using Eq. 1, while the black dotted lines show the position of the uncoupled IT and LSPR.

## Electron Energy-Loss Spectroscopy

To obtain further information and access to a wide range of plasmonic resonances, we performed on the same sample EELS measurements in a scanning transmission electron microscope (STEM).[59] STEM-EELS spectroscopy concurrently provides a high-angle annular dark-field (HAADF) image showing the topography of the structure, and a spectral image.[60] The EELS signal in the spectral image is almost directly proportional to the z-component of the electromagnetic local density of states (EMLDOS) at the position of the electron probe.[61] This means that spectrally, EELS gives a signature very close to extinction[62] while allowing to probe dark modes. This makes EEL spectroscopy a powerful experimental technique in plasmonics, allowing one to map surface plasmon resonances with nanoscale resolution[63-66], to study the plasmonic modes of coupled nanostructures[67] and to experimentally probe dispersion relations.[69,69] EELS has also been previously used to probe strong coupling between surface plasmons and excitons,[12,16] as well as between surface plasmons and phonons.[70]

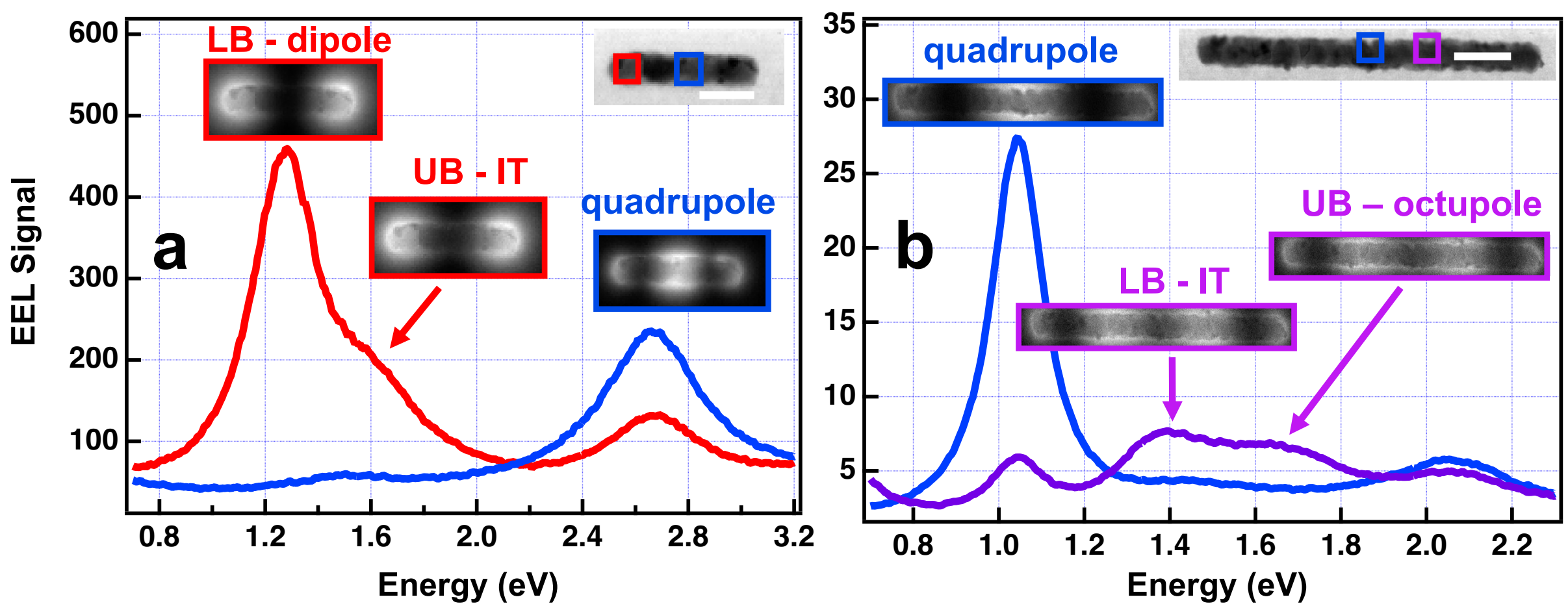

**Figure 3.** EEL spectra acquired at different locations on two Al nanorods of varying lengths: (**a**) 200 nm and (**b**) 580 nm. The locations are indicated through a color code in the upper right insets with the corresponding HAADF–STEM image. The scale bars are 75 nm and 100 nm, respectively. UB and LB corresponds to the two polaritonic branches (UB: upper branch; LB: lower energy branch).

**Figure 3** presents typical EEL spectra measured on two aluminum nanorods with lengths of 200 nm and 580 nm. The HAADF images of the nanorods are presented in the upper right insets. The spectra were extracted from the two square regions delineated on the HAADF images, representing a summation of the individual spectra (one per pixel) within the specified area. The EEL spectra exhibit distinct energy loss peaks at approximately 1.30 eV, 1.60 eV and 2.60 eV for the shorter nanorod and at approximately 1.05 eV, 1.40 eV, 1.70 eV, and 2.10 eV for the longer nanorod, depending on the location of measurement. The peaks corresponding to even and odd multipolar plasmonic resonances sustained by each Al nanorod (dipole $m = 1$; quadrupole $m = 2$; and subsequent modes of order $m$ up to 4 for the longer nanorod) are associated with their energy-filtered EEL maps in **Figure 3**, revealing the near-field distribution of the different modes. In the case where the IT peak is small with respect to the nearby plasmon, the maps are ambiguous as they are dominated by the intensity of the tail of the plasmon (**Figure 3a**, red line). When both are of similar intensity (**Figure 3b**, purple line), filtered EELS maps are representative of each excitation spatial distribution. Remarkably, the polaritonic excitations have the same spatial distribution, which is that of the (uncoupled) plasmonic mode of a given order. Such a behavior, where the coupled modes spatial distribution is mimicking the surface plasmon one, has been observed for plasmon-exciton[16] and plasmon-phonon strong coupling.[70]

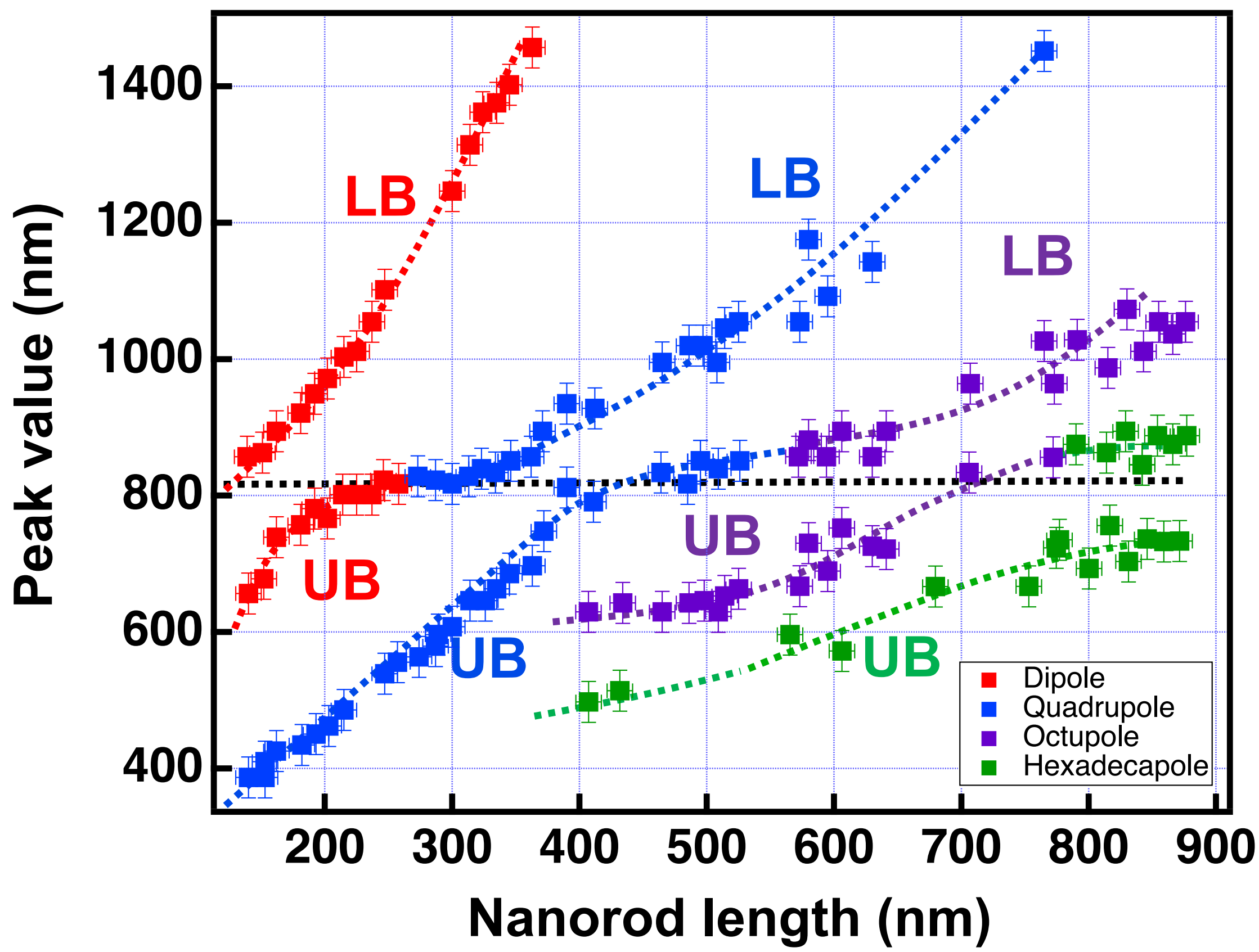

**Figure 4.** Peak wavelength positions extracted from the EEL spectra as a function of the nanorod length, for 72 different antennas. The colors are indicative of the order of the plasmonic resonance, as determined from the corresponding experimental energy-filtered maps. The colored dotted lines are a guide for the eyes. The horizontal dotted line shows the position of the uncoupled IT.

The resonance energies as a function of nanorod length were measured on 72 different antennas and are shown in **Figure 4**. For each data point, the full width at half maximum (FWHM) was extracted from the EEL spectra, and the resulting values are presented in Figure 4 as the error bars. Furthermore, the corresponding energy-filtered maps were employed to assign the order of the surface plasmon resonance to each point, with the results color-coded in **Figure 4**. The dispersion curve of the nanorods exhibits a splitting in three regions, which can be attributed to the coupling between IT and the three initial plasmonic modes. For each plasmonic mode two polaritonic branches are identified, one with a low energy (lower branch, LB) and one with a high energy (upper branch, UB) value.

The Rabi splitting was extracted from the EELS data for different nanorods using the previously described method, involving Eq. 1. The results are shown in **Figure 5** for the dipole, quadrupole and octupole plasmonic resonances and are compared with the findings from optical spectroscopy. A good agreement between the optical and electronic spectroscopies is observed, with comparable values for the Rabi splitting. It should be reminded that the quadrupole is a dark mode, which precludes its observation in the optical experiment. Interestingly, we observe that the high-order modes exhibit a reduced Rabi splitting energy in comparison to the dipole mode. A similar behavior has recently been reported in the context of strong coupling of the IT and surface plasmons in nickel.[25] In that reference, this effect was attributed to an increase in the plasmon linewidth of high-order multipolar plasmonic resonances in nickel. However, this is not the case for Al nanorods which exhibit relatively stable linewidths around the IT spectral position regardless of mode order.[54] Therefore, we hypothesize that the mechanisms responsible for the observed reduction in Rabi splitting in multipolar LSPR supported by Al nanorods can be attributed to their different mode volumes. This will be discussed below.

A thorough examination of **Figure 5** reveals a minor discrepancy between the extinction peak positions observed in EELS and those observed through optical spectroscopy. Specifically, the optical extinction appears to be slightly blue-shifted relative to the EELS data. A comparable discrepancy between EELS and optical spectroscopy has been previously reported by Husnik and co-workers,[71] an effect that could be related to the sample preparation prior to EELS experiments.

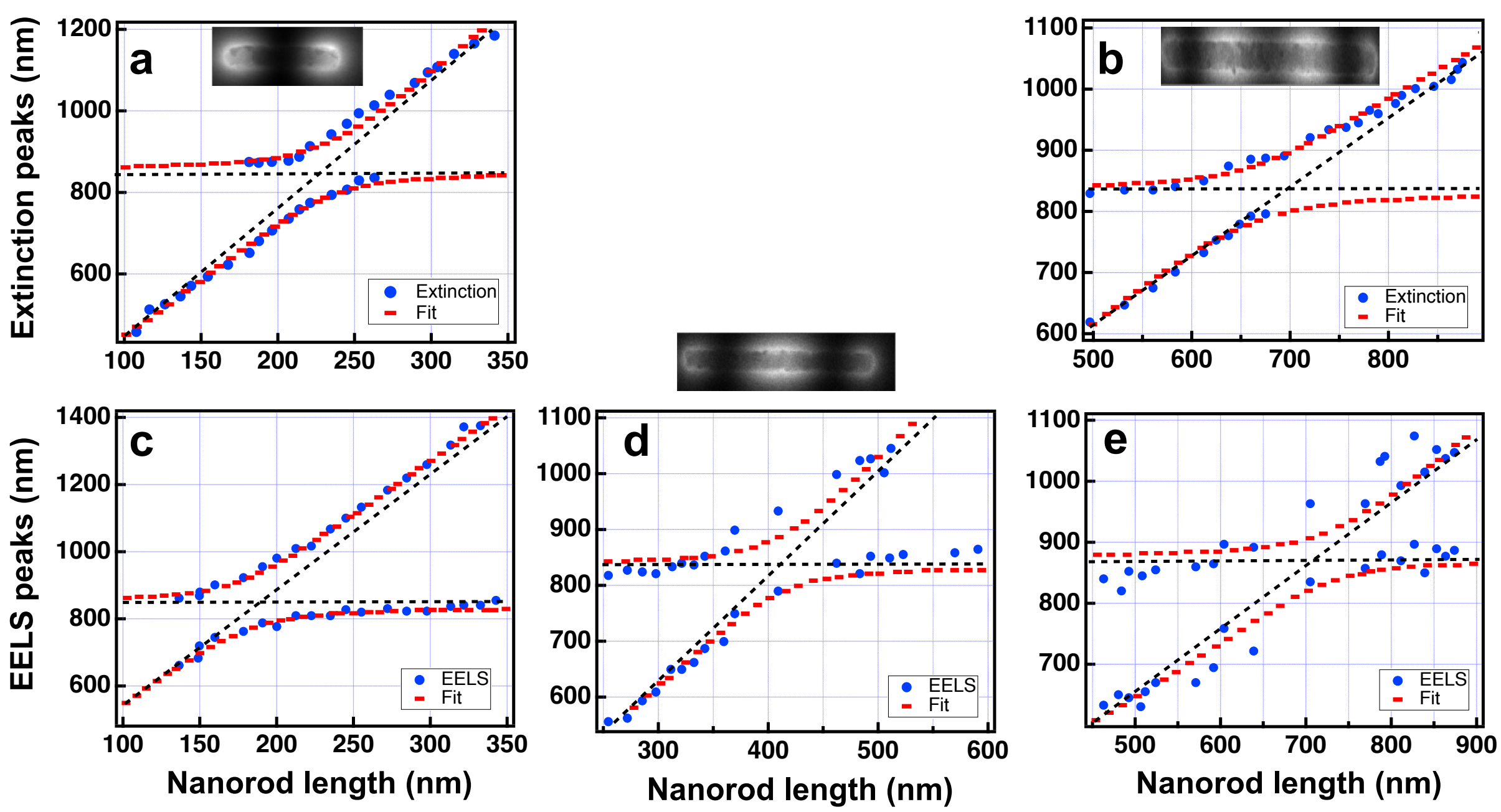

**Figure 5.** Dispersion diagrams (position of the extinction peaks vs. nanorod length) as retrieved from (a-b) optical extinction spectroscopy and (c-e) EELS. Each plot represents a single order of plasmonic resonance: dipole (a,c), quadrupole (d) and octupole (b,e). The blue dots correspond to the experimental data, the red solid line to a numerical fit using Eq 1. Upper insets: representative EEL maps of the corresponding modes.

**Table 1.** Summary of the different values of Rabi splitting energies (meV) calculated and measured in this study. The uncertainties given in the Table correspond to the standard deviations from the numerical fits (see Supplementary Information).

| | Dipole | Quadrupole | Octupole |
|---|---|---|---|
| FDTD | 229 ± 15 | *(dark)* | 153 ± 10 |
| Optical | 225 ± 10 | *(dark)* | 164 ± 8 |
| EELS | 250 ±10 | 172 ± 29 | 144 ± 33 |

**Analysis of the optical and EELS experiments**

The first question we want to address is whether strong coupling is being observed. It may seem surprising to observe strong coupling between two lossy oscillators such as a LSPR and the IT. However, these two resonances exhibit comparable FWHM (approximately 200 meV), a situation actually conducive to strong coupling, as pointed out in Ref. [72]. If we consider two oscillators characterized by complex eigenfrequencies $\hbar\widetilde{\omega_1} = \hbar\omega_1 + \frac{i}{2}\Gamma_1$ and $\hbar\widetilde{\omega_2} = \hbar\omega_2 + \frac{i}{2}\Gamma_2$, where the $\Gamma_i$ are the linewidths of the resonances, then the strong coupling condition is[58,73]

$$|\Gamma_1 - \Gamma_2| < 2g \qquad (2)$$

where $g = \hbar\Omega_R/2$ is the coupling strength. Thus, strong coupling is possible between the LSPR and the IT due to their similar linewidths. Furthermore, the strong coupling would be visible in the spectral domain if the Rabi splitting energy is larger than the individual linewidths of the resonance, i.e. $\hbar\Omega_R > \frac{1}{2}(\Gamma_1 + \Gamma_2)$, a condition that is verified in our case. Another topic worthy of discussion is the EELS distribution of the hybridized modes. As already emphasized, EELS filtered mapping does not yield an unambiguous image of the IT spatial distribution, as it is illustrated in **Figure 3a** for the dipole and in **Figure 3b** for the octupole. A striking feature is the similarity in the spatial distribution of the two modes (LSPR and IT), which aligns with the anticipated intensity distribution for a LSPR. Therefore, the IT, which is fundamentally a bulk excitation devoid of any local electromagnetic field enhancement, displays a non-uniform field distribution, which extends outside the nanoantenna, when hybridized with a LSPR. A similar behavior has indeed been predicted by FDTD simulations in the case of nickel nanodisks.[23] We emphasize that experimentally, it was not possible to observe the pure (uncoupled) IT of the nanorods using EELS. This is expected as the IT is supposed to have a vanishingly small intensity both in the bulk and near the surface of the metal in the general case.[59] When the IT energy becomes close to that of a LSPR, it can be discerned in the EEL spectra, manifesting as a shoulder in the LSPR peak. However, once it becomes observable, it is also coupled, and the observed mode is no longer a pure IT but a hybridized (polaritonic) mode with a LSPR component.

**Figure 6** provides further evidence of the polaritonic nature of the observed resonance. It compares EEL spectra recorded inside and outside two metallic nanorods for the dipole (**Figure 6a**) and quadrupole (**Figure 6b**) resonances. In each case, a signal in the spectral region corresponding to the IT is observed outside the nanorod, at locations where there is no metal. This demonstrates that the observed signal is not a pure IT (bulk excitation), but a hybridized mode that shares properties of both the IT and a LSPR. As the LSPR component extends outside the metal, the hybrid mode also extends outside the metal. As a reference, **Figure 6** also shows EEL spectra measured at a node of the LSPR resonance (gray squares in the insets of Figure 6). The corresponding spectra do not show any signal at the IT position, confirming that the IT cannot be observed outside the metal in the absence of a coupling to a LSPR.

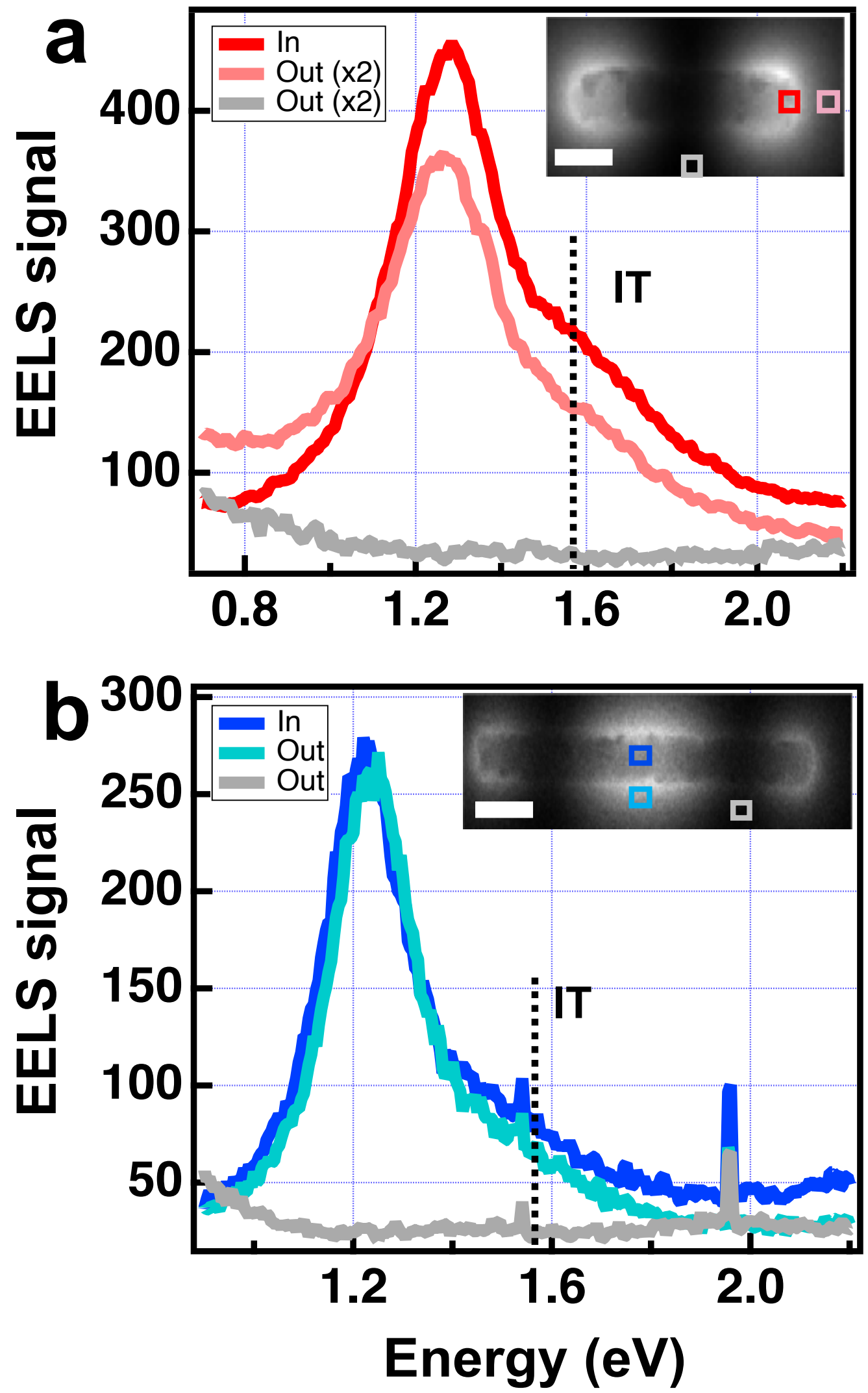

**Figure 6.** EEL spectra recorded inside and outside two Al nanorods with different length. **a** EEL spectra corresponding to the dipole resonance, recorded at one extremity of the rod with a length of 200 nm. The red line corresponds to a position within the rod, the pink line corresponds to a location outside the rod and the gray line corresponds to the position in a node of the plasmonic resonance (see the corresponding boxes in the inset). For enhanced visibility, the signals measured outside the nanorod have been multiplied by two. **b** EEL spectra corresponding to the quadrupole resonance, recorded at the center of the rod with length of 460 nm. The blue, light blue and gray lines correspond to positions within the rod, outside the rod and at a node of the quadrupolar

resonance, respectively. Insets: energy-filtered maps of the resonances. Scale bars are 50 and 80 nm, respectively.

Finally, we will examine how the Rabi energies evolve in conjunction with the different orders of the LSPR resonances, as illustrated in **Figure 5**. A summary of the Rabi energies, including those calculated by FDTD, is provided in **Table 1**. A discernible trend is observed, whereby the Rabi energy decreases with an increase in the order of the plasmonic mode. To elucidate this trend, we will examine the theoretical framework proposed by Eizner *et al.* to study strong coupling between excitons from J-aggregates and aluminum nanoantennas.[57] We recognize the inherent challenges in directly translating cQED models to nanophotonics,[74] however they offer a compelling physical interpretation of the observed trend in Rabi energies. In this model, the coupling (Rabi) energy is proportional to $\sqrt{\frac{N}{V}}$, where $N$ is the number of transition dipoles interacting (overlapping) with the plasmonic mode and $V$ is the mode volume associated with the LSPR. The definition of $V$ for small plasmonic resonators is more complex than for dielectric cavities due to the presence of both ohmic and radiative losses,[75,76] requiring the use of dedicated numerical methods to compute it.[77] In our case, two competing effects must be considered. On the one hand, for a given nanorod the plasmonic mode volume tends to decrease as the mode order increases (for instance, shifting from a dipole LSPR to a quadrupole LSPR). On the other hand, longer nanorods are required to observe the maximum Rabi splitting for higher order resonances (see **Figure 5**), implying larger mode volumes. Our experimental results suggest that $\sqrt{\frac{N}{V}}$ tends to decrease with the LSPR order. This hypothesis could be further substantiated through numerical computations of the plasmonic mode volumes at the IT frequency as a function of the nanorod length. However, this is beyond the scope of the present work.

**Physical structure of the extinction and hot electron generation at the interband transition**

The above results show that Al nanorods sustain hybrid modes, resulting from the strong coupling between the (multipolar) LSPR and the interband transition. These hybrid modes exhibit a broad spectral range spanning the visible and near-infrared regions. Now, we investigate the potential of these hybrid modes to generate hot electrons, employing a quantum formalism that enables us to calculate the efficiency of HE generation. This model incorporates Kreibig's approach[43] for the intraband hot electron generation as well as the nonlinear approach developed in Refs[45,48,78]. This nonlinear noniterative formalism has been implemented with COMSOL, allowing for the realistic modelling of the major mechanisms of dissipation and hot carrier generation, while being fully self-consistent, i.e., energy conserving.

For the Al nanorods the extinction cross-section can be divided into four terms:

$$\sigma_{ext} = \sigma_{Drude,bulk} + \sigma_S + \sigma_{interband} + \sigma_{scat} \qquad (3)$$

These terms correspond to the Drude dissipation inside the nanorod, the dissipation due to the generation of the intraband HEs at the surfaces (Kriebig's mechanism), the interband absorption in Al in the visible, and the scattering cross-section, respectively. At the same time, the COMSOL electromagnetic calculation, based on Maxwell's equations and the optical theorem, generates the extinction as a sum of two terms, $\sigma_{ext} = \sigma_{abs} + \sigma_{scat}$, where $\sigma_{abs}$ and $\sigma_{scat}$ are the absorption and scattering cross sections, respectively. Therefore, $\sigma_{abs} = \sigma_{Drude,bulk} + \sigma_S + \sigma_{interband}$, i.e., the absorption is composed of three terms corresponding to three mechanisms: the Drude dissipation (electron-phonon scattering), the intraband HE generation (surface-mediated electron scattering), and the interband HE generation. These processes are schematically shown in **Figure 7a**. The COMSOL model for the nanoantennas is shown in **Figure S2**. The Methods section gives more details of our formalism, which employs the following expression for the local dielectric function:

$$\varepsilon_{NR} = 1 + \Delta\varepsilon_{Drude} + \Delta\varepsilon_{s,HE-intraband} + \Delta\varepsilon_{interband,Lorentz} \quad (4)$$

where the Drude and Lorentz terms come from the bulk dielectric constant and $\Delta\varepsilon_{s,HE-intraband}$ describes the Kreibig's mechanism, i.e., the intraband HE generation.

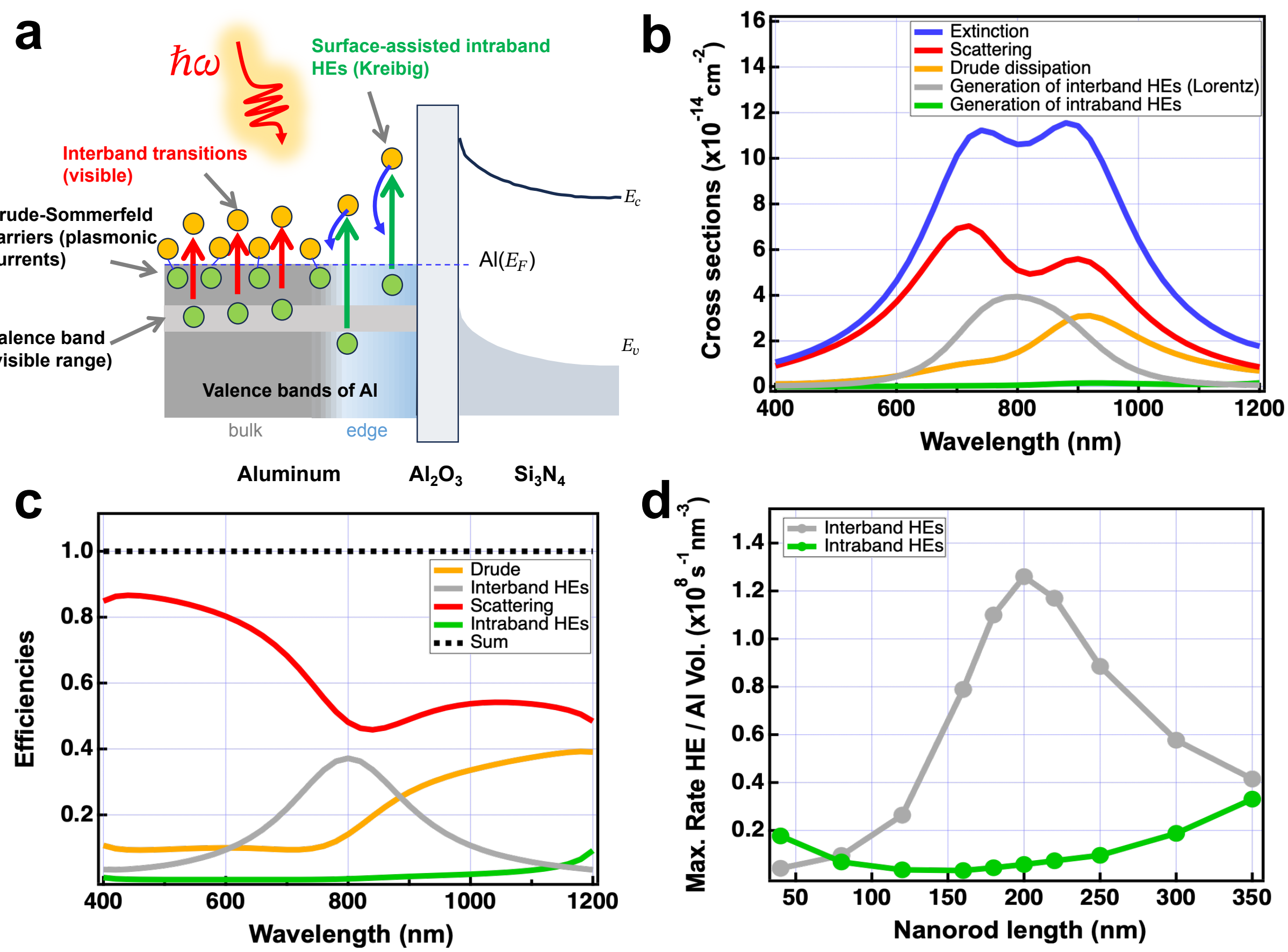

**Figure 7.** Hot electron generation in aluminum nanorods. **a** Schematic illustrating the different HE generation processes in an Al nanoantenna on top of a $Si_3N_4$ substrate upon illumination. The participating mechanisms are the Drude dissipation inside the nanorod, the dissipation due to the generation of the intraband HEs at the surfaces (Kriebig's mechanism, green arrows), the interband absorption in Al in the visible (red arrows), and the scattering cross-section. **b** Calculated absorption (Drude, interband, and intraband), scattering and extinction cross sections vs. wavelength. **c** Calculated efficiencies as a function of wavelength, showing the four terms contributing to the total extinction. We also include the sum (=1, dashed black line) which proves conservation of energy. Panels (**b**) and (**c**) were calculated for a L=200 nm Al nanoantenna (i.e. at the anti-crossing) via Comsol Multiphysics for the simulation conditions shown in Figure S2; for L=160 nm and L=250 nm see Figure S4. **d** Intraband and interband maximum HE generation rates for different nanorod lengths. The rates have been normalized by the corresponding nanorod's volume to ease the comparison.

**Figure 7b** shows the four contributions to the power dissipation separately. **Figure S3** already showed the total extinction calculated using COMSOL, and now we look at its components. What distinguishes an Al nanostructure from traditional noble metals, such as Au and Ag? Firstly, it is observed that the peak associated with the IT (depicted by the gray line in Figure 7b) manifests in the visible and near-infrared region (between 700 nm and 900 nm), whereas the interband excitations in Au and Ag are situated in the blue and UV regions. This indicates that the generation of interband HEs in Al may be used to harvest visible/near-IR photon quanta, which are plentiful in solar light, as opposed to UV photons. Secondly, the intraband dipolar plasmon can be tuned through the discrete interband transition, creating an interband-plasmon hybrid state, by varying the NR aspect ratio. Thirdly, the two hybrid resonances (i.e., the Rabi splitting in the spectrum) offer the potential to enhance the generation rates of the interband electrons and holes through the utilization of the strong dipolar plasmon resonance, which creates remarkable enhancement of the incident field inside the NR. This opportunity is not available in Au and Ag, as the plasmon resonances in the noble metals are typically in the green region or at longer wavelengths, while the interband transitions are situated in the blue and UV regions. Finally, it is notable that the energy distribution of the hot carriers generated via direct excitation of the interband transition in Al differs from that observed in noble metals. In gold, for instance, the IT promotes electrons from the d-band (which is well below the Fermi level) to the sp-band. This results in HE with an energy in the range of 0.5 – 1 eV above the Fermi level.[79] On the other hand, the hot holes generated in the d-band in gold have a high potential energy,[45] but with a low mean free path.[31] In contrast, Al can generate HE with higher energies.

These are distinctive properties of Al nanostructures that can be utilized for optical and energy-related applications. **Figure 7b** shows all contributions to the power dissipation $P = \sigma I_0$, for an incident flux $I_0$ = 2.5 × $10^7$ mW/$cm^2$. **Figure 7c** illustrates the efficiencies of the hot carriers and optical processes, which are defined as follows:[48]

$$\mathrm{Eff_{HE}} = P_{\mathrm{HE}}/P_{\mathrm{ext}} \quad (5)$$

$$\mathrm{Eff_{interband}} = P_{\mathrm{interband}}/P_{\mathrm{ext}} \quad (6)$$

$$\mathrm{Eff_{scat}} = P_{\mathrm{scat}}/P_{\mathrm{ext}} \quad (7)$$

$$\mathrm{Eff_{thermal}} = P_{\mathrm{thermal}}/P_{\mathrm{ext}} \quad (8)$$

where $P_{ext} = \sigma_{ext} I_0$ is the total power extracted from the incident flux. Conservation of energy dictates that $\mathrm{Eff_{HE}} + \mathrm{Eff_{interband}} + \mathrm{Eff_{Drude}} + \mathrm{Eff_{scat}} = 1$ and $\mathrm{Eff_{thermal}} + \mathrm{Eff_{scat}} = 1$, where $\mathrm{Eff_{thermal}}$ is the light-to-heat conversion efficiency, defined as $Eff_{thermal} = P_{thermal}/(P_{abs} + P_{scat})$; this parameter describes how much optical energy is converted into heat inside the plasmonic structure. The remaining efficiencies provide insight into the mechanisms by which the optical power is dissipated within the plasmonic system. As anticipated, the intraband HE generation is weak in the nanorods with large sizes, i.e., $Eff_{HE} \ll 1$ (green curve in **Figure 7c**). This is due to the fact that HE generation (Kreibig's mechanism) is a surface scattering effect. The most noteworthy result shown in **Figure 7c** is the remarkably high efficiency of photon conversion into interband electron-hole pairs. For the peak values the efficiency is 40%. In previous works[47,48] we observed that the Ag-related interband transitions in Ag-coated gold nanorods show unusually high efficiencies. However, such transitions in silver occur in the UV region, whereas the Al NRs calculated here have a very high efficiency of light-to-hot carrier conversion in the visible region. The latter mechanism is particularly attractive for solar energy harvesting. Finally, $Eff_{scat}$ and $Eff_{thermal}$ are anti-correlated, since the scattering process typically decreases with increasing wavelength. Therefore $Eff_{thermal}(\lambda)$ is an overall increasing function of λ.

The rates of generation of intraband and interband HEs are defined as $Rate_{\text{HE}} = P_{\text{HE}}/\hbar\omega$ and $Rate_{\text{interband}} = P_{\text{interband}}/\hbar\omega$. These rates are wavelength dependent, and representative spectra of the rates are shown in **Figure S4** for two different rod lengths: 160 nm and 250 nm, which correspond to two different hybridization configurations. In Figure S4a-c (shortest rod) the dipole plasmon is located on the short wavelength side of the IT, whereas in Figure S4d-f the dipole is on the long wavelength side of the IT. In each case, the maximum interband HE generation efficiency is approximately 40%. In contrast, the interband HE generation rate is higher for the longer rod (see Figures S4c and f). Additionally, the two spectra exhibit notable differences, with the spectrum for the longest rod displaying a slight red shift compared to the 160 nm rod. In order to demonstrate that the mode hybridization between the IT and the surface plasmon is favorable for the generation of HE, we have computed the *maximum* HE generation rate (i.e., the rate at the peak wavelength) for different rod lengths. As the number of electrons generated is also dependent on the volume of the metal, the rates were normalized with the volume of the corresponding nanorod. This procedure ensures that any observed effect is not merely a trivial consequence of the size of the electron reservoir. The results are presented in **Figure 7d**. A clear maximum in the normalized interband HE generation rate is observed for a length of 200 nm, which coincides with the length at which mode hybridization is observed (see, e.g., **Figure 2a** and **S3**). In other words, when the interband is hybridized (strongly coupled) with a surface plasmon resonance, an anomalously high rate of HE is generated in the metal.

To better understand the HE generation, **Figure 8** presents maps illustrating the spatial distribution of the hot electrons, for the $L = 200$ nm nanorod. These maps show the locations where electrons with energies exceeding the Schottky barrier of the metal-substrate interface are generated. The maps have been computed for three distinct wavelengths corresponding to the lower and upper polaritons and the center of the gap (see **Figure 8a**). The hot electron maps are compared with the electric field enhancement $|E|/|E_0|$ maps at three different heights inside the nanorod (**Figure 8b**). **Figure 8c** shows the results of our computations. For all three wavelengths ($\lambda = 723, 809$ and 901 nm), we observe that the electric field distribution is strongly dependent on the height $z$. In particular, at $z = 0$ (i.e. at the interface between the Al nanoantennas and the $Si_3N_4$ substrate) intense and highly localized hot spots are observed at the rounded corners of the rod, which corresponds to the spatial distribution of the dipole plasmon. As can be seen in the maps of HE generation (rightmost panel in **Figure 8c**), the HEs are generated at these electric field hot spots. This indicates that the physical mechanism responsible for the enhancement of HE generation is an electromagnetic process: the interband-surface plasmon hybridization creates regions where the optical intensity is locally enhanced at photon energies corresponding to the IT, thereby efficiently promoting electrons towards higher energy bands. Additionally, the highest generation rate of hot electrons appears to occur at the bottom of the nanorod. In contrast, the hot electron maps for the top surface (not shown here) show no hot spots and no noticeable rate variations. This information is crucial for applications in photodetection, as the generated HE could be transferred to the substrate and subsequently collected, depending on their mean free path.

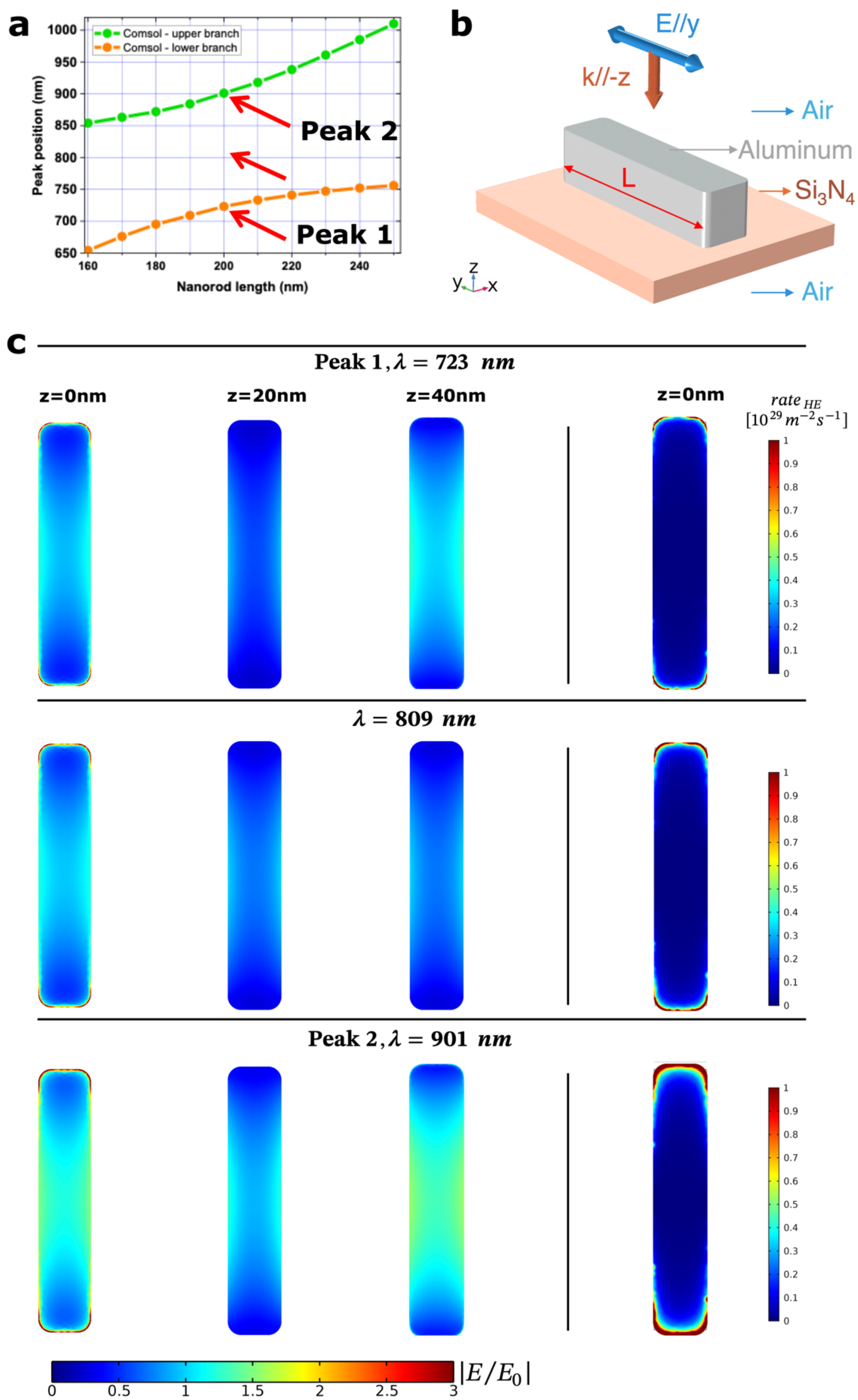

a
Comsol - upper branch
Comsol - lower branch
Peak 2
Peak 1
Peak position (nm)
Nanorod length (nm)
b
E//y
k//-z
Air
Aluminum
L
$Si_3N_4$
Air
c
Peak 1, $\lambda = 723$ nm
z=0nm
z=20nm
z=40nm
z=0nm
$rate_{HE}$ $[10^{29} m^{-2}s^{-1}]$
$\lambda = 809$ nm
Peak 2, $\lambda = 901$ nm
$|E/E_0|$

**Figure 8.** **a** Evolution of the wavelength of the extinction peak maxima with length using COMSOL Multiphysics. The three red arrows indicate the two peaks and the dip for the *L*=200 nm Al nanoantenna, for which we calculate E-field enhancement maps in panel (**c**) below. **b** Schematic of the simulated Al nanoantenna. The dimensions are the same as those used for the FDTD calculations of Figure 1, i.e. for the Al nanoantenna the width $t_1$=40 nm, the height $t_1$=40 nm, the vertical edge rounding *r*=10nm. The length $L$ varies from 150 to 250 nm, for the $Si_3N_4$ substrate the thickness $t_2$=15 nm. The surrounding media are as shown in the figure, with $\varepsilon(Si_3N_4)$=4, $\varepsilon_{air}$=1, and $\varepsilon_{Al}$ given by a Drude-Lorentz fit. Linearly $E_y$ polarized light is incident along the $k_z$ direction, with an intensity of $2.5\times10^7$ $mW/cm^2$. **c** Electric field enhancement maps showing planar sections of the *L* = 200 nm Al nanoantenna, for the three wavelengths shown in panel (**a**). The sections are for *z* = 0, 20 and 40 nm, so at the bottom (interface with the substrate) at the middle and at the top of the Al nanoantenna, respectively. The right-hand columns of the maps are the surface HE rates at the bottom of the rod (*z*=0). The hot electron maps for the top side of the rod show no hot spots and no noticeable rate variations.

## CONCLUSION

Using simple individual aluminum nanostructures (nanorods), we have experimentally evidenced a phenomenon of self-hybridization of the plasmonic resonances sustained by the nanoantenna. This phenomenon is due to the strong coupling of a localized plasmonic mode with interband transitions occurring within the metal. The strong coupling was investigated over a wide spectral range using both extinction spectroscopy and electron energy loss spectroscopy, which revealed the characteristic anti-crossing behavior of the initial resonances. The large number of samples studied allowed us to collect a wealth of experimental data, both in the spectral regions where strong coupling occurs and away from them (see **Figure 4**). Such systematic measurements are critical to avoid possible selection biases, as recently underlined by Thomas & Barnes.[80] Moreover, energy-filtered maps from the EELS experiments provided what is, to the best of our knowledge, the first image of a hybridized LSPR-IT mode. From this image, we demonstrated that the hybrid resonances exhibited characteristics of both the IT and the LSPR, thereby substantiating their polaritonic nature. This self-strong coupling phenomenon can occur in any material where the IT is spectrally localized, thus enabling the creation of "cavity-free" polaritonic states via self-coupling.[81] The strong coupling was observed for multiple orders of the plasmon resonance, and it was demonstrated that the associated Rabi energy can be tuned by simply changing the order of the LSPR involved. Such a result is likely due to a change in the corresponding plasmonic mode volumes.

The second key result is the prediction of anomalously high rates of hot electron generation associated with hybridization. Although the generation of interband HE in metals is well documented, the hybridization of the IT with a localized surface plasmon was not addressed in the literature so far. In this novel scheme, a localized surface plasmon creates strong electromagnetic hot spots inside the metal, dramatically enhancing the overall light-assisted generation of interband HE. The volume-normalized generation rate passes through a clear maximum for nanorods whose size corresponds to the strong coupling condition. The associated interband HE efficiency is about 40%. Another important result of our study is the development of a model for the decomposition of the total extinction cross-section into four components, namely scattering, Drude-absorption, IT-absorption, and surface-assisted dissipation. Our theoretical framework is fully self-consistent, locally conserving charge and energy.

Finally, we stress that this phenomenon is triggered by red and near-infrared light, a spectral range that is particularly well suited for solar light-based applications, such as hot carrier-based photodetection. For conventional plasmonic metals such as Au and Ag, the hybridization regime is not observed because the interband transitions are not spectrally isolated. It is also worth noting that in addition to its optical properties, the appeal of Al-based systems lies in their low material cost and well-developed lithographic technology. In that respect, the Al system is unique and worthy of further investigation.

## METHODS

### FDTD calculations

To model the optical properties of the Al nanoantennas, we used a commercial software (Ansys Lumerical FDTD). The modelled structure consists of a single aluminum nanorod of variable length, with a width w = 40 nm and a height h = 40 nm. The structure is placed on top of a 15 nm thick dielectric substrate with a real dielectric index $\varepsilon = 4$, corresponding to the average value of silicon nitride for visible and near-infrared wavelengths. The complex permittivity of aluminum was obtained from the CRC Handbook of Chemistry and Physics,[55] which reproduced its Drude-Lorentz profile and captured the absorption peak centered at $\omega_{IT}$ = 820 nm attributed to the IT. To model single nanorods, we used a Total-Field Scattered-Field (TFSF) source, which allows the separation of incident and scattered fields. Perfectly Matched Layers (PMLs) absorbing boundary conditions were used for all directions of space. A non-uniform mesh was employed to model the nanorod, with a minimum mesh step of 0.25 nm. A linearly polarized light source was considered, with its polarization direction aligned with the long axis of the nanorod. The extinction efficiency is defined as the ratio between the extinction cross-section and the geometrical area (top surface) of the nanostructure.

**COMSOL modelling of hot electron generation**

To model the optical properties and the hot electron generation of the Al nanoantennas, we employed a commercial software (COMSOL Multiphysics® software, with the RF module). The geometry of the structure was identical to that described for FDTD, as detailed in Figure S2. The complex permittivity of bulk aluminum was obtained from the experimental Palik's data.[82] Then, the data was fitted to a Drude-Lorentz model in order to obtain the Drude and the interband (Lorentz) contributions to the absorption. The system was built as a single scatterer on a substrate (see Ref. [83]) where a plane TE-polarized electromagnetic wave was incident on an Al nanoantenna on silicon nitride. The absorption, scattering and extinction cross-sections of the nanoantenna were computed for a normal incidence.

*Scatterer on substrate* means that the model initially calculated a background field based on the plane wave incident on the substrate only (i.e. in the absence of the Al nanorod). This was then used to determine the total field with the nanoantenna present. Briefly, the system is composed of two primary components: the substrate-scatterer-media subdomain, and an additional subdomain surrounding it consisting of a perfectly matched layer (PML) box, which accounts for infinite media, thereby creating a single scatterer. The main features of the model can be found in [83]. The specific parameters for the simulation are as follows: the substrate-scatterer-media subdomain has dimensions of $600 \times 600$ nm$^2$ along $x$ and $y$, and 800 nm in height (along $z$). The mesh setting is "Free Tetrahedral", and the mesh size is set to the predetermined "Finer" setting in COMSOL (Maximum element size=44 nm, minimum element size=3.2 nm, maximum element growth

rate=1.4, Curvature factor=0.4, Resolution of narrow regions=0.7). Within this subdomain, the type of boundary conditions used along $x$ and $y$ in the absence of the scatterer was set as "Continuity". The PML exhibits a thickness of 80 nm in all directions, and the previously described mesh size "Finer". The settings "Distribution" (10 points for each *x*-*y*-*z* directions, 5 points for the *z*-direction of the $Si_3N_4$ substrate, 5 points for the PML's corner), "Mapped" and "Swept" were used to generate cubic domains for the entire PML.

As explained above (Eq. 4), there are three contributions to the local dielectric function: the Drude dissipation, the interband HEs (Lorentz) and the intraband HE. In conjunction with the scattering, this results in a total of four contributions to the extinction (Eq. 5-8) and correspondingly, to four contributions to the efficiency of carriers. The contributions of the HEs were included in accordance with the self-consistent method developed in Refs. [45,48,78]. The dielectric constant $\varepsilon_{Al} = Drude - Lorentz$ is effectively modified by an effective quantum parameter, $\gamma_S$, which is a plasmonic broadening due to the surface-mediated electron scattering, as follows:

$$\varepsilon_{Al}(\omega) = 1 + \frac{G_0\omega_0^2}{\omega_0^2-\omega^2-i\Gamma\omega} - \frac{\omega_p^2}{\omega\left(\omega+i(\gamma_D+\gamma_S)\right)}. \quad (9)$$

Each contribution can be defined as follows:

$$\varepsilon_{Drude-Surface,HE}(\omega) = 1 - \frac{\omega_p^2}{\omega\left(\omega+i(\gamma_D+\gamma_S)\right)} = \varepsilon_D + \varepsilon_{Surface,HE}, \quad (10a)$$

$$\varepsilon_D = 1 - \frac{\omega_p^2}{\omega\left(\omega+i(\gamma_D)\right)} \quad (10b)$$

$$\varepsilon_{Surface,HE} = -\frac{\omega_p^2}{\omega\left(\omega+i(\gamma_D+\gamma_S)\right)} + \frac{\omega_p^2}{\omega\left(\omega+i(\gamma_D)\right)} \quad (10c)$$

$$\varepsilon_{Lorentz} = \frac{G_0\omega_0^2}{\omega_0^2-\omega^2-i\Gamma\omega} \quad (10d)$$

The contribution of each heat term was then calculated as:

$$Q_D = \frac{\varepsilon_0}{2\omega} \mathrm{Im}[\varepsilon_D(\omega)] E_\omega \cdot E_\omega^*, \qquad (11a)$$

$$Q_{Surface,HE} = \frac{\varepsilon_0}{2\omega} \mathrm{Im}\left[\varepsilon_{Surface,HE}(\omega)\right] E_\omega \cdot E_\omega^*, \qquad (11b)$$

$$Q_{Lorentz} = \frac{\varepsilon_0}{2\omega} \mathrm{Im}[\varepsilon_{Lorentz}(\omega)] E_\omega \cdot E_\omega^*. \qquad (11c)$$

The broadening term $\gamma_S$ contained in both $\varepsilon_{Surface,HE}$ and $Q_{Surface,HE}$, was calculated with an iterative numerical process defined by

$$\gamma_S = \frac{3}{4} v_F \frac{\int_S |E_{normal}(\theta,\varphi)|^2 \mathrm{ds}}{\int_{NC} E_\omega \cdot E_\omega^* \mathrm{d}V} = \frac{3}{4} v_F \frac{S}{V} \frac{F_S(\gamma_S)}{F_v(\gamma_S)} \qquad (12)$$

where $v_F = 2.03 \times 10^6 m/s$ is the Fermi velocity for aluminum[84] and

$$F_S(\gamma_S) = \frac{1}{S} \int_S |E_{normal}(\theta, \varphi)|^2 \mathrm{d}s, \qquad (13)$$

$$F_V(\gamma_S) = \frac{1}{V} \int_{NC} E_\omega \cdot E_\omega^* \mathrm{d}V, \qquad (14)$$

where *S* and *V* are the surface and volume, respectively, of the nanorod under study and $E_{normal}$ is the electric field normal to the surface inside the metal. For full details on the derivation see Ref. [45]. Lastly, the local maps of HE generation shown in Figure 8 were calculated as

$$Rate_{HE}(\boldsymbol{r}) = \frac{1}{4} \times \frac{2}{\pi^2} \times \frac{e^2 E_F^2}{\hbar} \frac{(\hbar\omega - \Delta E_{bar})}{(\hbar\omega)^4} |E_{normal}(\theta, \varphi)|^2 \qquad (15)$$

where $E_F = 11.7$ eV is the Fermi energy for aluminum[84] and is $\Delta E_{bar}$ the injection threshold energy (Schottky barrier of the aluminum-substrate interface). In our computations we used $\Delta E_{bar} = 0.69$ eV.[85]

**Sample fabrication**

The nanorod arrays were fabricated using electron beam lithography (EBL) in a scanning electron microscope equipped with a field emission gun (eLine; Raith). First, a layer of poly(methyl methacrylate) resist with a thickness of 150 nm was spin-coated on a TEM–EELS compatible substrate (15 nm-thick $Si_3N_4$ square membranes from Ted Pella, Inc). Subsequently, the resist was insulated by the electron beam via the EBL system. The imprinted patterns were then developed for 60 s in a 1:3 MIBK:IPA solution at room temperature. Finally, a 40 nm-thick layer of Al was deposited on the sample using thermal evaporation (ME300; Plassys), followed by lift-off in acetone.

**Extinction spectroscopy**

For extinction measurements we employed a custom-built optical setup. The illumination was generated by a broadband laser-driven light source (EQ-99X; Energetic), which produced a stable signal between 170 and 2500 nm. The light from the source was routed via fiber optics to a silver parabolic mirror for beam collimation, and then to a linear polarizer aligned with the nanorods' long axis. Subsequently, the beam was weakly focused on the sample with the aid of a lens. The transmitted light was collected by a long working-distance microscope objective, with a numerical aperture of 0.65 (Mitutoyo ×50). Ultimately, the light was injected directly into the slit of the spectrometer (ISOPLANE-160; Princeton Instruments) in order to maximize the signal-to-noise ratio. Altogether, this system defined a collection area of approximately 25×25 $\mu m^2$, which is smaller than the lithographed area.

**Electron energy loss spectroscopy and imaging**

EEL spectroscopy was conducted using a monochromated NION Hermes200 microscope, which was equipped with an IRIS spectrometer. The acceleration voltage was set to 100 keV and the energy resolution was in the range 24-30 meV. The typical spectral images were 200 pixels per 200 pixels, with the dimensions adjusted to align with the given structure. All EEL spectra have been post-processed using an alignment procedure based on the zero-loss peak position.

ASSOCIATED CONTENT

**Supporting Information**.

Supplementary Information (PDF): This file contains four Supplementary Figures and one Supplementary Method (detailed fitting procedure).

Figure S1: FDTD calculation of the strong coupling between the IT and the octupole

Figure S2: COMSOL model of the Al nanorods

Figure S3: COMSOL calculations of the extinction efficiency of Al nanorods

Figure S4: Interband and intraband HE generation for two different nanorod lengths (L=160 nm and L=250 nm).

AUTHOR INFORMATION

**Corresponding Author**

*E-mail: jerome.martin@utt.fr ; davy.gerard@utt.fr

**Author Contributions**

The manuscript was written through contributions of all authors. All authors have given approval to the final version of the manuscript.

**Funding Sources**

French National Agency for Research (grants ANR-20-CE30-0033, ANR-10-EQPX-50, ANR-10-EQPX-50); European Union (grants 823717 ESTEEM3 and 101017720 EBEAM); Région Grand Est.

ACKNOWLEDGMENT

Samples were fabricated on the Nanomat platform (www.nanomat.eu) with support from the French RENATECH+ network. TS acknowledges support from the Région Grand Est. This work has been done within the framework of the Graduate School NANO-PHOT (grant ANR-10-EQPX-50). This work received support from the National Agency for Research under grant QUENOT (ANR-20-CE30-0033) and the program of future investment TEMPOS-CHROMATEM (ANR-10-EQPX-50). This project also received funding from the European Union’s Horizon 2020 Research and Innovation Program under grants 823717 (ESTEEM3) and 101017720 (EBEAM). The authors acknowledge financial support from the CNRS-CEA “METSA” French network (FR CNRS 3507) on the platform LPS-STEM. This work was also partially supported by the National Social Science Fund of China (22&ZD046) and the “111 Project” (B20030).

REFERENCES

1. Haroche, S.; Raimond, J.-M. *Exploring the quantum: atoms, cavities, and photons*; Oxford University Press: Oxford, 2006.

2. Garcia-Vidal, F. J.; Ciuti, C.; Ebbesen, T. W. Manipulating matter by strong coupling to vacuum fields. Science 2021, 373, eabd0336.

3. Novotny, L. Strong coupling, energy splitting, and level crossings: A classical perspective. Am. J. Phys. 2010, 78, 1199–1202.

4. Lidzey, D. G.; Bradley, D.; Skolnick, M.; Virgili, T.; Walker, S.; Whittaker, D. Strong exciton–photon coupling in an organic semiconductor microcavity. Nature 1998, 395, 53–55.

5. Törmä, P.; Barnes, W. L. Strong coupling between surface plasmon polaritons and emitters: a review. Rep. Prog. Phys. 2014, 78, 013901.

6. Bellessa, J.; Bonnand, C.; Plenet, J.-C.; Mugnier, J. Strong coupling between surface plasmons and excitons in an organic semiconductor. Phys. Rev. Lett. 2004, 93, 036404.

7. Symonds, C.; Bonnand, C.; Plenet, J.; Brehier, A.; Parashkov, R.; Lauret, J.-S.; Deleporte, E.; Bellessa, J. Particularities of surface plasmon–exciton strong coupling with large Rabi splitting. New J. Phys. 2008, 10, 065017.

8. Zengin, G.; Wersäll, M.; Nilsson, S.; Antosiewicz, T. J.; Käll, M.; Shegai, T. Realizing strong light-matter interactions between single-nanoparticle plasmons and molecular excitons at ambient conditions. Phys. Rev. Lett. 2015, 114, 157401.

9. Hakala, T.; Toppari, J.; Kuzyk, A.; Pettersson, M.; Tikkanen, H.; Kunttu, H.; Törmä, P. Vacuum Rabi splitting and strong-coupling dynamics for surface-plasmon polaritons and rhodamine 6G molecules. Phys. Rev. Lett. 2009, 103, 053602.

10. Gomez, D. E.; Vernon, K. C.; Mulvaney, P.; Davis, T. J. Surface plasmon mediated strong exciton-photon coupling in semiconductor nanocrystals. Nano Lett. 2010, 10, 274–278.

11. Baudrion, A.-L.; Perron, A.; Veltri, A.; Bouhelier, A.; Adam, P.-M.; Bachelot, R. Reversible strong coupling in silver nanoparticle arrays using photochromic molecules. Nano Lett. 2013, 13, 282–286.

12. Wei, J.; Jiang, N.; Xu, J.; Bai, X.; Liu, J. Strong coupling between ZnO excitons and localized surface plasmons of silver nanoparticles studied by STEM-EELS. Nano Lett. 2015, 15, 5926–5931.

13. Wen, J.; Wang, H.; Wang, W.; Deng, Z.; Zhuang, C.; Zhang, Y.; Liu, F.; She, J.; Chen, J.; Chen, H.; Deng, S.; Xu, N. Room-temperature strong light–matter interaction with active control in single plasmonic nanorod coupled with two-dimensional atomic crystals. Nano Lett. 2017, 17, 4689–4697.

14. Zheng, D.; Zhang, S.; Deng, Q.; Kang, M.; Nordlander, P.; Xu, H. Manipulating coherent plasmon–exciton interaction in a single silver nanorod on monolayer $WSe_2$. Nano Lett. 2017, 17, 3809–3814.

15. Cuadra, J.; Baranov, D. G.; Wersall, M.; Verre, R.; Antosiewicz, T. J.; Shegai, T. Observation of tunable charged exciton polaritons in hybrid monolayer $WS_2$ – plasmonic nanoantenna system. Nano Lett. 2018, 18, 1777–1785.

16. Yankovich, A. B.; Munkhbat, B.; Baranov, D. G.; Cuadra, J.; Olsen, E.; Lourenço-Martins, H.; Tizei, L. H.; Kociak, M.; Olsson, E.; Shegai, T. Visualizing spatial variations of plasmon–exciton polaritons at the nanoscale using electron microscopy. Nano Lett. 2019, 19, 8171–8181.

17. Li, S.; Ai, R.; Chui, K. K.; Fang, Y.; Lai, Y.; Zhuo, X.; Shao, L.; Wang, J.; Lin, H. Q. Routing the Exciton Emissions of $WS_2$ Monolayer with the High-Order Plasmon Modes of Ag Nanorods. Nano Lett. 2023, 23, 4183-4190.

18. Chikkaraddy, R.; De Nijs, B.; Benz, F.; Barrow, S. J.; Scherman, O. A.; Rosta, E.; Demetriadou, A.; Fox, P.; Hess, O.; Baumberg, J. J. Single-molecule strong coupling at room temperature in plasmonic nanocavities. Nature 2016, 535, 127–130.

19. Baranov, D. G.; Wersall, M.; Cuadra, J.; Antosiewicz, T. J.; Shegai, T. Novel nanostructures and materials for strong light–matter interactions. ACS Photonics 2018, 5, 24–42.

20. Pakizeh, T. Optical absorption of plasmonic nanoparticles in presence of a local interband transition. J. Phys. Chem. C 2011, 115, 21826–21831.

21. Schwind, M.; Kasemo, B.; Zoric, I. Localized and propagating plasmons in metal films with nanoholes. Nano Lett. 2013, 13, 1743–1750.

22. Lecarme, O.; Sun, Q.; Ueno, K.; Misawa, H. Robust and versatile light absorption at near-infrared wavelengths by plasmonic aluminum nanorods. ACS Photonics 2014, 1, 538–546.

23. Pirzadeh, Z.; Pakizeh, T.; Miljkovic, V.; Langhammer, C.; Dmitriev, A. Plasmon–interband coupling in nickel nanoantennas. ACS Photonics 2014, 1, 158–162.

24. Schuermans, S.; Maurer, T.; Martin, J.; Moussy, J.-B.; Plain, J. Plasmon/interband transitions coupling in the UV from large scale nanostructured Ni films. Opt. Mater. Express 2017, 7, 1787–1793.

25. Assadillayev, A.; Faniayeu, I.; Dmitriev, A.; Raza, S. Nanoscale engineering of optical strong coupling inside metals. Adv. Opt. Mater. 2023, 11, 2201971.

26. Lourenço-Martins, H.; Das, P.; Tizei, L. H.; Weil, R.; Kociak, M. Self-hybridization within non-Hermitian localized plasmonic systems. Nat. Phys. 2018, 14, 360-364.

27. Tserkezis, C.; Stamatopoulou, P. E.; Wolff, C.; Mortensen, N. A. Self-hybridisation between interband transitions and Mie modes in dielectric nanoparticles. Nanophotonics 2024, 13, 2513-2522.

28. Kittel, C. *Quantum Theory of Solids*, 2$^{nd}$ Revised Edition; John Wiley & Sons: New York, 1991.

29. Manjavacas, A.; Liu, J. G.; Kulkarni, V.; Nordlander, P. Plasmon-induced hot carriers in metallic nanoparticles. ACS Nano 2014, 8, 7630-7638.

30. Lee, S. A.; Link, S. Chemical interface damping of surface plasmon resonances. Acc. Chem. Res. 2021, 54, 1950-1960.

31. Khurgin, J.; Bykov, A. Y.; Zayats, A. V. Hot-electron dynamics in plasmonic nanostructures: fundamentals, applications and overlooked aspects. eLight 2024, 4(1), 15.

32. Schirato, A.; Maiuri, M.; Cerullo, G.; Della Valle, G. Ultrafast hot electron dynamics in plasmonic nanostructures: experiments, modelling, design. Nanophotonics 2023, 12, 1-28.

33. Hartland, G. V.; Besteiro, L. V.; Johns, P.; Govorov, A. O. What's so hot about electrons in metal nanoparticles? ACS Energy Lett. 2017, 2, 1641–1653.

34. Kim, M.; Lin, M.; Son, J.; Xu, H.; Nam, J.-M. Hot-electron-mediated photochemical reactions: principles, recent advances, and challenges. Adv. Opt. Mater. 2017, 5, 1700004.

35. Zhang, Y.; He, S.; Guo, W.; Hu, Y.; Huang, J.; Mulcahy, J. R.; Wei, W. D. Surface-plasmon-driven hot electron photochemistry. Chem. Rev. 2017, 118, 2927–2954.

36. Wang, F.; Melosh, N. A. Plasmonic energy collection through hot carrier extraction. Nano Lett. 2011, 11, 5426-5430.

37. Tang, H.; Chen, C.-J.; Huang, Z.; Bright, J.; Meng, G.; Liu, R.-S.; Wu, N. Plasmonic hot electrons for sensing, photodetection, and solar energy applications: A perspective. J. Chem. Phys. 2020, 152, 220901.

38. Chalabi, H.; Schoen, D.; Brongersma, M. L. Hot-electron photodetection with a plasmonic nanostripe antenna. Nano Lett. 2014, 14, 1374-1380.

39. Li, W.; Coppens, Z. J.; Besteiro, L. V.; Wang, W.; Govorov, A. O.; Valentine, J. Circularly polarized light detection with hot electrons in chiral plasmonic metamaterials. Nature Commun. 2015, 6(1), 8379.

40. Wu, S.; Hogan, N.; Sheldon, M. Hot electron emission in plasmonic thermionic converters. ACS Energy Lett. 2019, 4, 2508-2513.

41. Zhao, J.; Zhong, Y.; Zhang, L.; Sui, L.; Wu, G.; Zhang, J.; Han, K.; Zhang, Q.; Yuan, K.; Yang, X. Relaxation Channels of Two Types of Hot Carriers in Gold Nanostructures. Nano Lett. 2024, 24, 15340-15347.

42. Maier, S. A. *Plasmonics: fundamentals and applications*; Springer: New York, 2007.

43. Kreibig, U.; Vollmer, M. *Optical properties of metal clusters*; Springer Series in Materials Science vol. 25. Springer: Berlin, 1995.

44. Besteiro, L. V.; Yu, P.; Wang, Z.; Holleitner, A. W.; Hartland, G. V.; Wiederrecht, G. P.; Govorov, A. O. The fast and the furious: Ultrafast hot electrons in plasmonic metastructures. Size and structure matter. Nano Today 2019, 27, 120-145.

45. Santiago, E. Y.; Besteiro, L. V.; Kong, X.-T.; Correa-Duarte, M. A.; Wang, Z.; Govorov, A. O. Efficiency of hot-electron generation in plasmonic nanocrystals with complex shapes: surface-induced scattering, hot spots, and interband transitions. ACS Photonics 2020, 7, 2807–2824.

46. Linic, S.; Chavez, S.; Elias, R. Flow and extraction of energy and charge carriers in hybrid plasmonic nanostructures. Nat. Mater. 2021, 20, 916-924.

47. Negrin-Montecelo, Y.; Comesaña-Hermo, M.; Khorashad, L. K.; Sousa-Castillo, A.; Wang, Z.; Perez-Lorenzo, M.; Liedl, T.; Govorov, A. O.; Correa-Duarte, M. A. Photophysical effects behind the efficiency of hot electron injection in plasmon-assisted catalysis: the joint role of morphology and composition. ACS Energy Lett. 2019, 5, 395–402.

48. Muravitskaya, A.; Movsesyan, A.; Avalos-Ovando, O.; Bahamondes Lorca, V. A.; Correa-Duarte, M. A.; Besteiro, L. V.; Liedl, T.; Yu, P.; Wang, Z.; Markovich, G.; Govorov, A.O. Hot Electrons and Electromagnetic Effects in the Broadband Au, Ag, and Ag–Au Nanocrystals: The UV, visible, and NIR Plasmons. ACS Photonics 2023, 11, 68–84.

49. Gérard, D.; Gray, S. K. Aluminium plasmonics. J. Phys. D: Appl. Phys. 2015, 48, 184001.

50. Ehrenreich, H.; Philipp, H. R.; Segall, B. Optical properties of aluminum. Phys. Rev. 1963, 132, 1918.

51. Ashcroft, N. W.; Sturm, K. Interband absorption and the optical properties of polyvalent metals. Phys. Rev. B 1971, 3, 1898.

52. Simon, T.; Li, X.; Martin, J.; Khlopin, D.; Stéphan, O.; Kociak, M.; Gérard, D. Aluminum Cayley trees as scalable, broadband, multiresonant optical antennas. Proc. Natl. Acad. Sci. U.S.A. 2022, 119, e2116833119.

53. Knight, M. W.; Liu, L.; Wang, Y.; Brown, L.; Mukherjee, S.; King, N. S.; Everitt, H. O.; Nordlander, P.; Halas, N. J. Aluminum plasmonic nanoantennas. Nano Lett. 2012, 12, 6000–6004.

54. Martin, J.; Kociak, M.; Mahfoud, Z.; Proust, J.; Gérard, D.; Plain, J. High-resolution imaging and spectroscopy of multipolar plasmonic resonances in aluminum nanoantennas. Nano Lett. 2014, 14, 5517–5523.

55. Lide, D. R. (Ed.) CRC handbook of chemistry and physics, 88$^{th}$ ed.; CRC press, 2007.

56. Novotny, L.; Hecht, B. *Principles of nano-optics,* 2$^{nd}$ edition; Cambridge University Press: Cambridge, 2012.

57. Eizner, E.; Avayu, O.; Ditcovski, R.; Ellenbogen, T. Aluminum nanoantenna complexes for strong coupling between excitons and localized surface plasmons. Nano Lett. 2015, 15, 6215–6221.

58. Benisty, H.; Greffet, J. J.; Lalanne, P. *Introduction to nanophotonics*; Oxford University Press: Oxford, 2022.

59. Egerton, R. F. *Electron energy-loss spectroscopy in the electron microscope,* 3rd edition; Springer: New York, 2011.

60. Kociak, M.; Stéphan, O. Mapping plasmons at the nanometer scale in an electron microscope. Chem. Soc. Rev. 2014, 43, 3865–3883.

61. Garcia De Abajo, F.; Kociak, M. Probing the photonic local density of states with electron energy loss spectroscopy. Phys. Rev. Lett. 2008, 100, 106804.

62. Losquin, A.; Zagonel, L. F.; Myroshnychenko, V.; Rodríguez-González, B.; Tencé, M.; Scarabelli, L.; Förstner, J.; Liz-Marzán, L.M.; García de Abajo, F.J.; Stéphan, O.; Kociak, M. Unveiling nanometer scale extinction and scattering phenomena through combined electron energy loss spectroscopy and cathodoluminescence measurements. Nano Lett. 2015, 15, 1229-1237.

63. Nelayah, J.; Kociak, M.; Stéphan, O.; Garcia de Abajo, F. J.; Tencé, M.; Henrard, L.; Taverna, D.; Pastoriza-Santos, I.; Liz-Marzán, L. M.; Colliex, C. Mapping surface plasmons on a single metallic nanoparticle. Nat. Phys. 2007, 3, 348–353.

64. Rossouw, D.; Couillard, M.; Vickery, J.; Kumacheva, E.; Botton, G. Multipolar plasmonic resonances in silver nanowire antennas imaged with a subnanometer electron probe. Nano Lett. 2011, 11, 1499–1504.

65. Koh, A. L.; Fernández-Domínguez, A. I.; McComb, D. W.; Maier, S. A.; Yang, J. K. High-resolution mapping of electron-beam-excited plasmon modes in lithographically defined gold nanostructures. Nano Lett. 2011, 11, 1323-1330.

66. Bosman, M.; Ye, E.; Tan, S. F.; Nijhuis, C. A.; Yang, J. K.; Marty, R.; Mlayah, A.; Arbouet, A.; Girard, C.; Han, M. Y. Surface plasmon damping quantified with an electron nanoprobe. Sci. Rep. 2013, 3, 1312.

67. Duan, H.; Fernández-Domínguez, A. I.; Bosman, M.; Maier, S. A.; Yang, J. K. Nanoplasmonics: classical down to the nanometer scale. Nano Lett. 2012, 12, 1683-1689.

68. Campos, A.; Arbouet, A.; Martin, J.; Gérard, D.; Proust, J.; Plain, J.; Kociak, M. Plasmonic breathing and edge modes in aluminum nanotriangles. ACS Photonics 2017, 4, 1257–1263.

69. Saito, H.; Lourenço-Martins, H.; Bonnet, N.; Li, X.; Lovejoy, T. C.; Dellby, N.; Stéphan, O.; Kociak, M.; Tizei, L. H. G. Emergence of point defect states in a plasmonic crystal. Phys. Rev. B 2019, 100, 245402.

70. Tizei, L. H.; Mkhitaryan, V.; Lourenço-Martins, H.; Scarabelli, L.; Watanabe, K.; Taniguchi, T.; Tencé, M.; Blazit, J.-D.; Li, X.; Gloter, A.; Zobelli, A. ; Schmidt, F.-P.; Liz-Marzán, L. M.; García de Abajo, F.J.; Stéphan, O.; Kociak, M. Tailored nanoscale plasmon-enhanced vibrational electron spectroscopy. Nano Lett. 2020, 20, 2973-2979.

71. Husnik, M.; von Cube, F.; Irsen, S.; Linden, S.; Niegemann, J.; Busch, K.; Wegener, M. Comparison of electron energy-loss and quantitative optical spectroscopy on individual optical gold antennas. Nanophotonics 2013, 2, 241-245.

72. Faucheaux, J. A.; Fu, J.; Jain, P. K. Unified theoretical framework for realizing diverse regimes of strong coupling between plasmons and electronic transitions. J. Phys. Chem. C 2014, 118, 2710–2717.

73. Chen, M., Marguet, S., Issa, A., Jradi, S., Couteau, C., Fiorini-Debuisschert, C., Douillard, L., Soppera, O., Ge, D., Plain, J., Zhou, X., Dang, C., Béal, J., Kostcheev, S., Déturche, R., Xu, T., Wei ; B., Bachelot, R. Approaches for Positioning the Active Medium in Hybrid Nanoplasmonics. Focus on Plasmon-Assisted Photopolymerization. ACS Photonics 2024, 11, 10, 3933–3953.

74. Tserkezis, C.; Fernández-Domínguez, A. I.; Gonçalves, P.; Todisco, F.; Cox, J. D.; Busch, K.; Stenger, N.; Bozhevolnyi, S. I.; Mortensen, N. A.; Wolff, C. On the applicability of quantum-optical concepts in strong-coupling nanophotonics. Rep. Prog. Phys. 2020, 83, 082401.

75. Derom, S.; Vincent, R.; Bouhelier, A.; Colas des Francs, G. Resonance quality, radiative/ohmic losses and modal volume of Mie plasmons. Europhys. Lett. 2012, 98, 47008.

76. Koenderink, A. F. On the use of Purcell factors for plasmon antennas. Opt. Lett. 2010, 35, 4208-4210.

77. Lalanne, P.; Yan, W.; Vynck, K.; Sauvan, C.; Hugonin, J. P. Light interaction with photonic and plasmonic resonances. Laser Photonics Rev. 2018, 12, 1700113.

78. Movsesyan, A.; Santiago, E. Y.; Burger, S.; Correa-Duarte, M. A.; Besteiro, L. V.; Wang, Z.; Govorov, A. O. Plasmonic Nanocrystals with Complex Shapes for Photocatalysis and Growth: Contrasting Anisotropic Hot-Electron Generation with the Photothermal Effect. Adv. Opt. Mater. 2022, 10, 2102663.

79. Goswami, A., Kim, A. S., & Cai, W. Exploring the synergy between hot-electron dynamics and active plasmonics: A perspective. J. Appl. Phys. 2024, 136, 100901.

80. Thomas, P. A.; Barnes, W. L. Selection bias in strong coupling experiments. J. Phys. Chem. Lett. 2024, 15, 1708-1710.

81. Canales, A.; Baranov, D. G.; Antosiewicz, T. J.; Shegai, T. Abundance of cavity-free polaritonic states in resonant materials and nanostructures. J. Chem. Phys. 2021, 154, 024701.

82. Palik, E. D. (Ed.). *Handbook of optical constants of solids*. Academic Press: San Diego, 1985.

83. *Scatterer on substrate*, COMSOL. https://www.comsol.com/model/scatterer-on-substrate-14699 (accessed May 22, 2025).

84. Ashcroft, N.; Mermin, N. *Solid State Physics*, 1st edition; Brooks/Cole: Belmont, 1976.

85. Yu, A.Y.C.; and Mead C.A. Characteristics of Aluminum-Silicon Schottky Barrier Diode. Solid-State Electron. 1970, 13, 97-104.

# Supplementary Information for: *Plasmon-interband hybridization and anomalous production of hot electrons in aluminum nanoantennas*

Jérôme Martin,*,† Oscar Avalos-Ovando,¶ Thomas Simon,† Gabriel Arditi,‡ Florian Lamaze,† Julien Proust,† Luiz Tizei,‡ Zhiming Wang,[1,2] Mathieu Kociak,‡ Alexander O. Govorov,¶ Odile Stéphan,‡ and Davy Gérard*,†

†Lumière, nanomatériaux, nanotechnologies (L2n), UMR CNRS 7076, Université de Technologie de Troyes, Troyes 10004, France

‡Laboratoire de Physique des Solides, Bâtiment 510, UMR CNRS 8502, Université Paris Saclay, Orsay 91400, France

¶Department of Physics and Astronomy and Nanoscale and Quantum Phenomena Institute, Ohio University, Athens, Ohio 45701, United States

1 Institute of Fundamental and Frontier Sciences, University of Electronic Science and Technology of China, Chengdu 611731, China

2 Shimmer Center, Tianfu Jiangxi Laboratory, Chengdu 641419, China

*E-mail: jerome.martin@utt.fr ; davy.gerard@utt.fr

**CONTENTS**

This Supplementary Information file contains four Supplementary Figures (Figures S1, S2, S3 and S4) and a Supplementary Method.

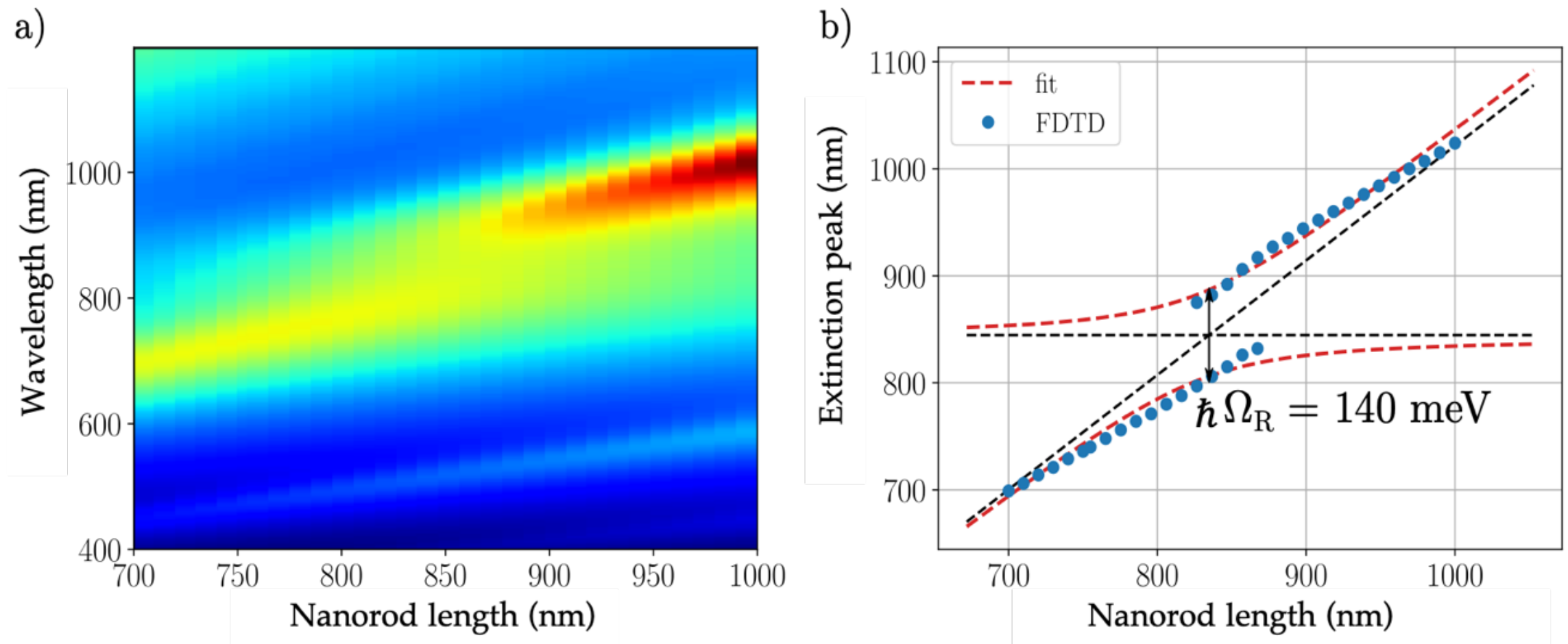

**Figure S1** Evolution of the extinction of an aluminum nanorod with its length in the case of the octupole mode calculated by FDTD, highlighting the impact of interband transitions around 800 nm. a) Extinction efficiency $Q_{ext}$ as a function of rod length and wavelength, and b) evolution of the extinction peak frequency with length, where the red dashed lines correspond to a numerical fit of the data.

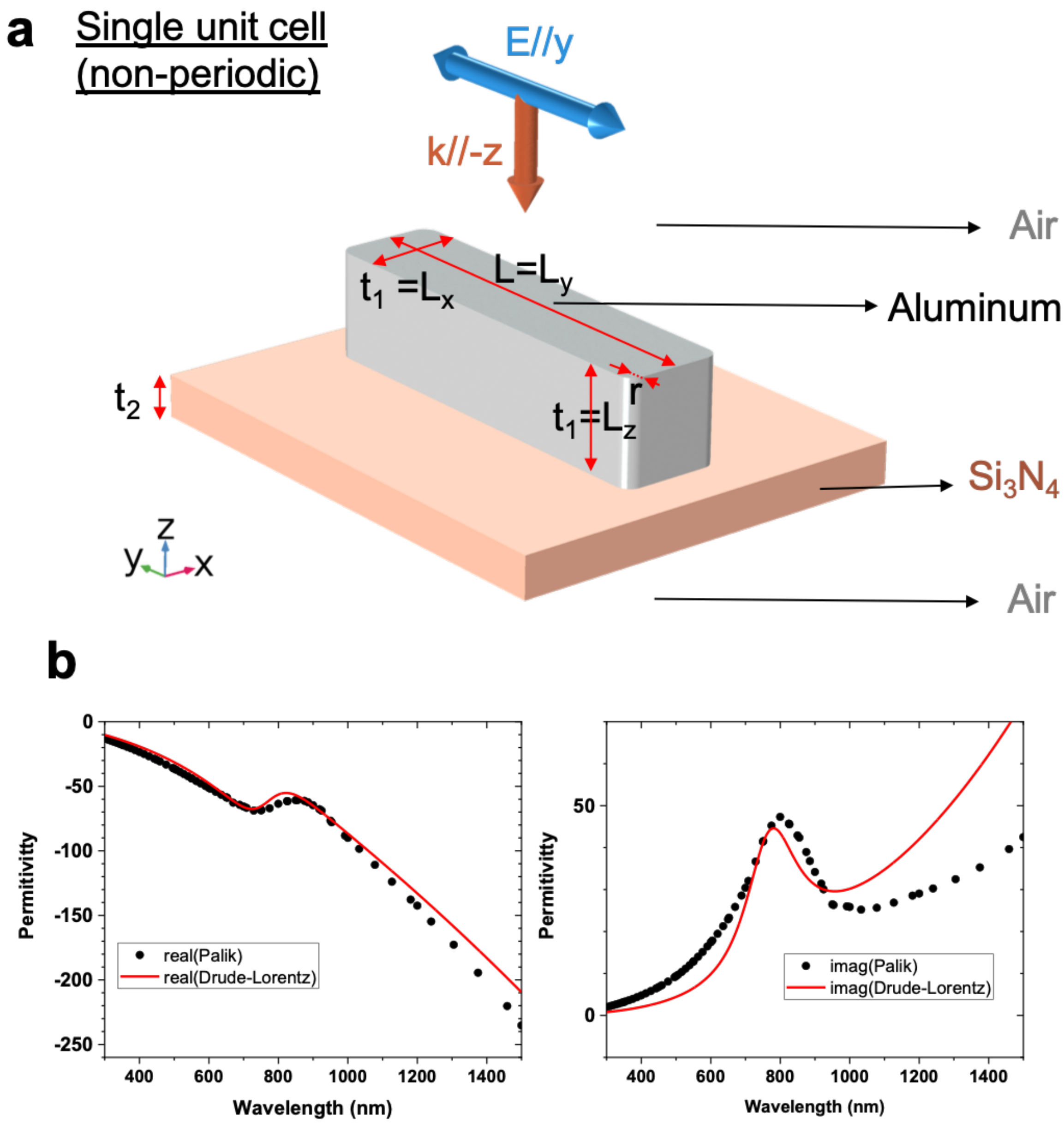

**Figure S2**. **a** Schematics of the Al nanoantenna simulated with finite elements via Comsol Multiphysics. The dimensions are the same used for the FDTD calculations of Figure 1, i.e. for the Al nanoantenna's width $t_1$=40 nm, height $t_1$=40 nm, vertical edge rounding r=10nm, and length L varies from 150 to 250 nm; for the $Si_3N_4$ substrate the thickness is $t_2$=15 nm. The media is as explained in the figure, with $\varepsilon_{Si3N4} = 4$, $\varepsilon_{air} = 1$, and $\varepsilon_{Al} = Drude - Lorentz$ fit. Linearly E//y polarized light is incident along the k//-z direction, with an intensity of $2.5 \times 10^7$ mW/cm$^2$. **b** Permittivity used in the Comsol Multiphysics simulations: Real part (left) and imaginary part (right) of the $\varepsilon_{Al} = Drude - Lorentz$ fit (solid red line) compared to the experimental available data of Palik's (black symbols).

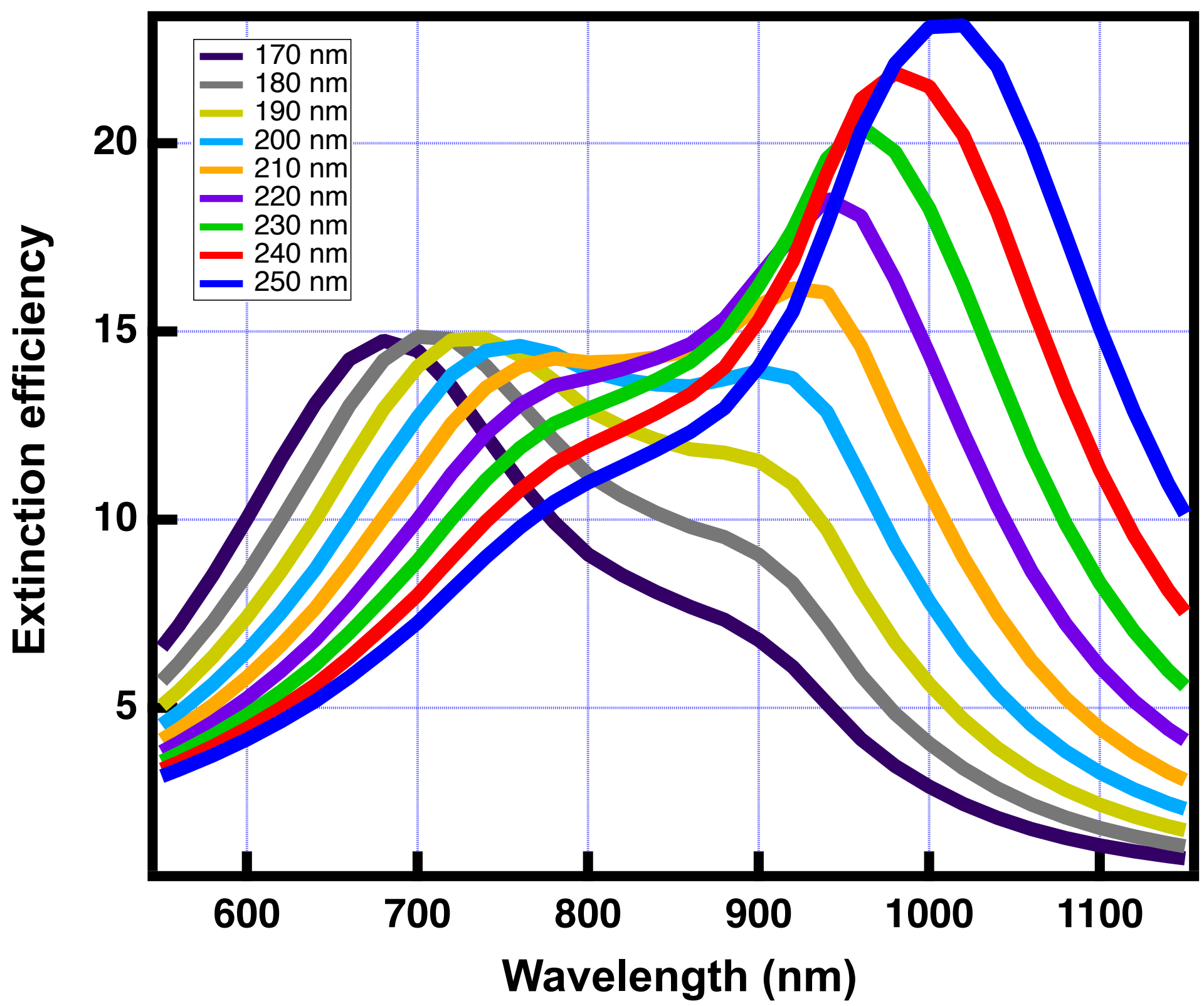

**Figure S3**. Extinction efficiency vs. wavelength for different nanorod lengths computed using the COMSOL model from Figure S2. The slight discrepancy in the peak maximum position between these results and those from Lumerical (Figure 1) is due to the different permittivities employed in the modelling of aluminum: the CRC Handbook [1] from Lumerical, and Palik [2] for COMSOL (with a correction to incorporate surface-related effects, see Methods).

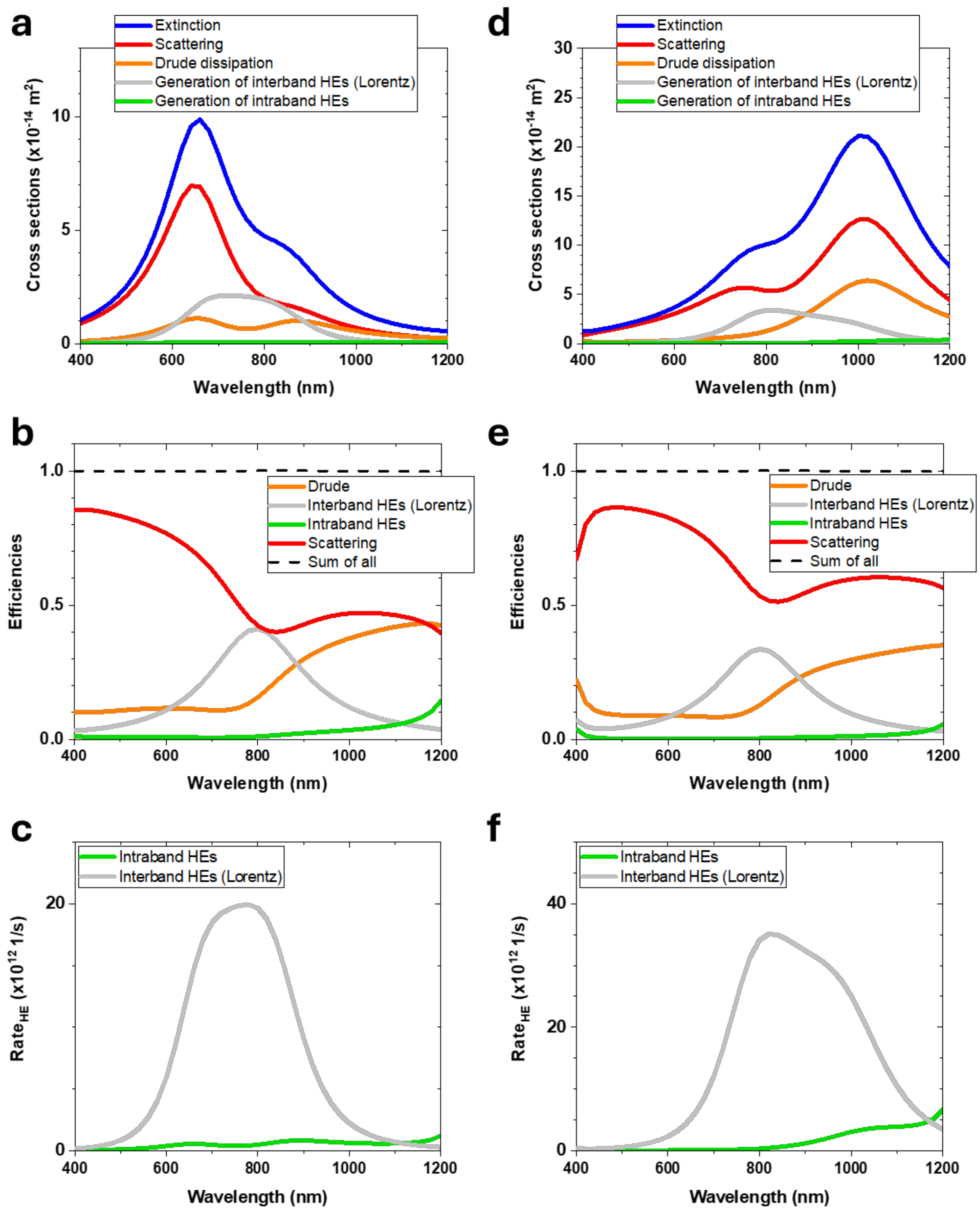

**Figure S4**. Panels (**a**)-(**c**) for a L=160 nm, and panels (**d**)-(**f**) for a L=250 nm Al nanoantenna. (**a**) and (**d**) are the absorption (Drude, interband, and intraband), scattering and extinction cross sections. (**b**) and (**e**) are the efficiencies of the four terms contributing to the total extinction, where we also include the sum (=1, dashed black line) which proves conservation of energy. (**c**) and (**f**) are the rates of intraband- and interband-HEs generation.

## Supplementary Method: fitting procedure

In the classical model of strong coupling we use the following equation to compute the energies $E_{\pm}$ of the upper and lower branches [3] :

$$E_{\pm} = \frac{1}{2}\left(\hbar\omega_{IT} + \hbar\omega_{SP} \pm \sqrt{\hbar^2\Omega_R^2 + (\hbar\omega_{IT} - \hbar\omega_{SP})^2}\right) \quad \text{(S1)}$$

where $\omega_{IT}$ is the angular frequency of the interband transitions (ITs) and $\omega_{SP}$ the angular frequency of the plasmonic modes *uncoupled* to ITs.

The energies of the uncoupled plasmonic modes can be calculated considering a "pure" Drude aluminum in FDTD calculations as illustrated in Figure S5.

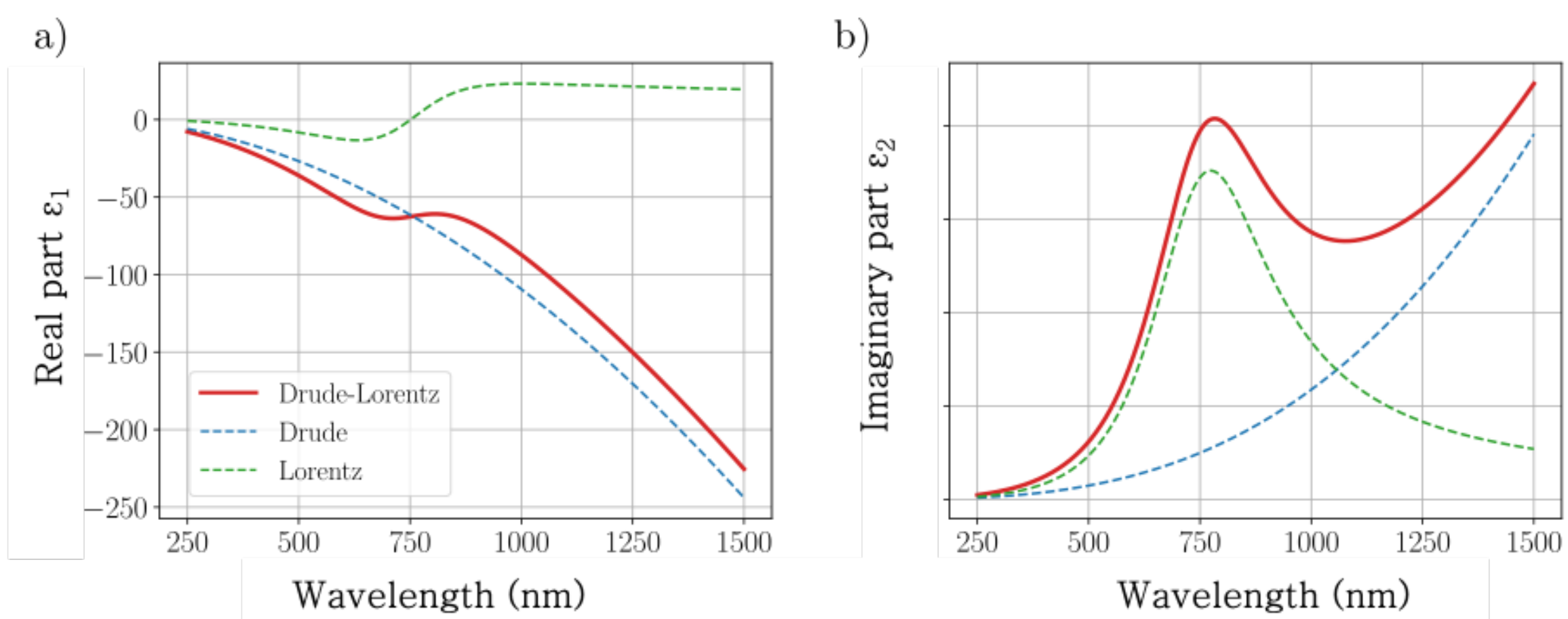

**Figure S5.** Dielectric permittivity of aluminum in the Drude-Lorentz model. a) Real part and b) imaginary part. The model coefficients are taken from Lecarme *et al.* [4]

The impact of using whether a Drude or a Drude-Lorentz function to model the permittivity of aluminum is illustrated in Figure S6 for the dipolar mode. Whereas the Drude-Lorentz model captures the impact of the IT, a pure Drude metal shows only the LSPR, whose peak wavelength increases when the rod length increases.

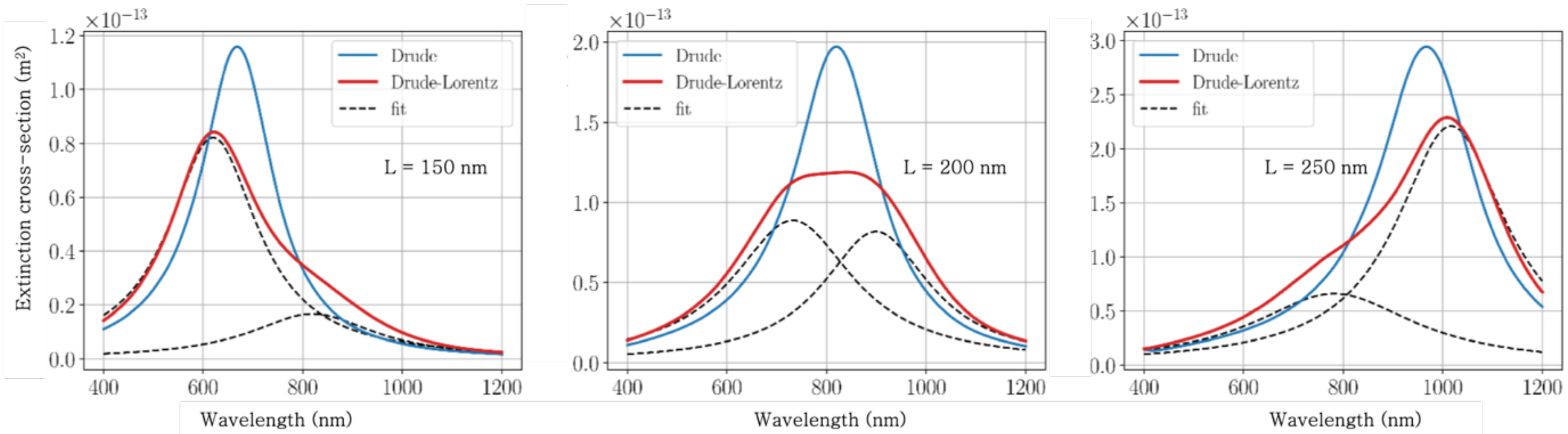

**Figure S6.** Impact of interband transitions on the dipolar plasmonic mode sustained by Al nanoantennas using FDTD simulations, for three different nanorod lengths. The red solid lines correspond to a calculation where the permittivity of Al is modelled with a Drude-Lorentz function, the blue solid lines to a Drude modeling, and the black dashed lines to a fit of the Drude-Lorentz calculation evidencing the two peaks.

In order to fit our numerical and experimental data with Equation (S1), a continuous dispersion relation for the uncoupled plasmonic modes is required. We assume that the LSPR energy follows the following quadratic dispersion:

$$\hbar\omega_{lsp} = ak^2 + bk + c \qquad \text{(S2)}$$

*a, b* and *c* being the fitting parameters given in $eV.m^{-2}$, $eV.m^{-1}$ and eV, respectively.

The wavevector is linked to the nanorod length assuming the relation $k = \frac{\pi}{L}n$, where $L$ is the length of the nanorod and $n$ is an integer corresponding to the order of the plasmonic mode [5]. As illustrated in Figure S7 for the dipolar plasmonic resonances, the dispersion relation (Eq. S2) fits very well with calculated uncoupled plasmonic resonances, confirming the validity of our approximation.

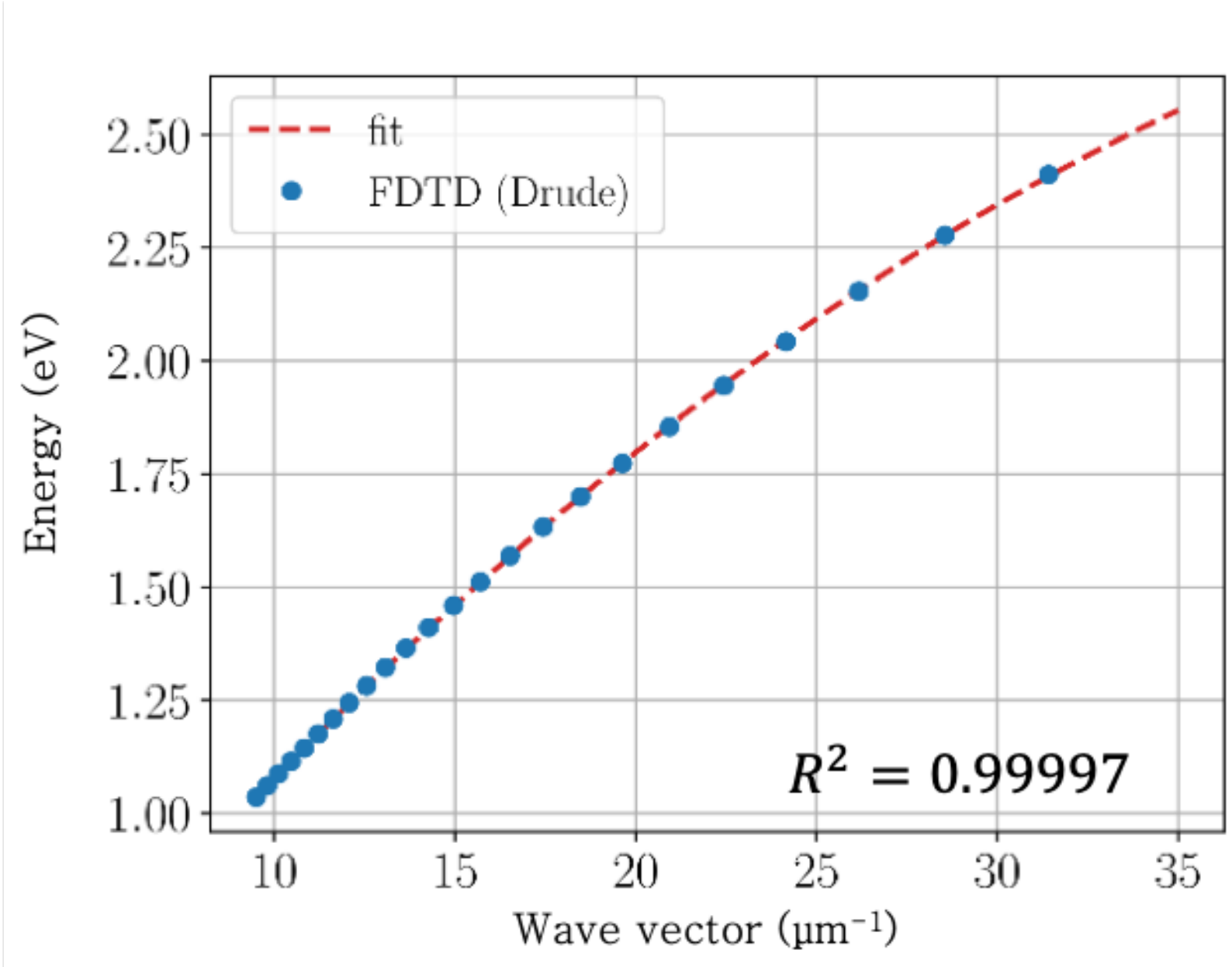

**Figure S7.** Calculated energy of plasmonic dipolar resonances sustained by a Drude-like Al nanoantenna ITs as a function of the wavevector. Fitting parameters: a = -1065 eV.m$^{-2}$, b = 115.2 eV.m$^{-1}$, c = -2962 eV.

Finally, we inject Equation (S2) in Equation (S1) to obtain the following relation:

$$E_{\pm} = \frac{1}{2}\left( \hbar\omega_{IT} + \left(a\frac{\pi^2 n^2}{L^2} + b\frac{\pi\, n}{L} + c\right) \pm \sqrt{\hbar^2\Omega_R^2 + \left(\hbar\omega_{IT} - \left(a\frac{\pi^2 n^2}{L^2} + b\frac{\pi\, n}{L} + c\right)\right)^2} \right) \quad \text{(S3)}$$

In the manuscript, Equation (S3) was used to fit the experimental data by adjusting the five parameters a, b, c, $\omega_{IT}$ and $\Omega_R$. A representative example is given in Figure S8 for EELS data corresponding to the dipolar plasmonic resonances.

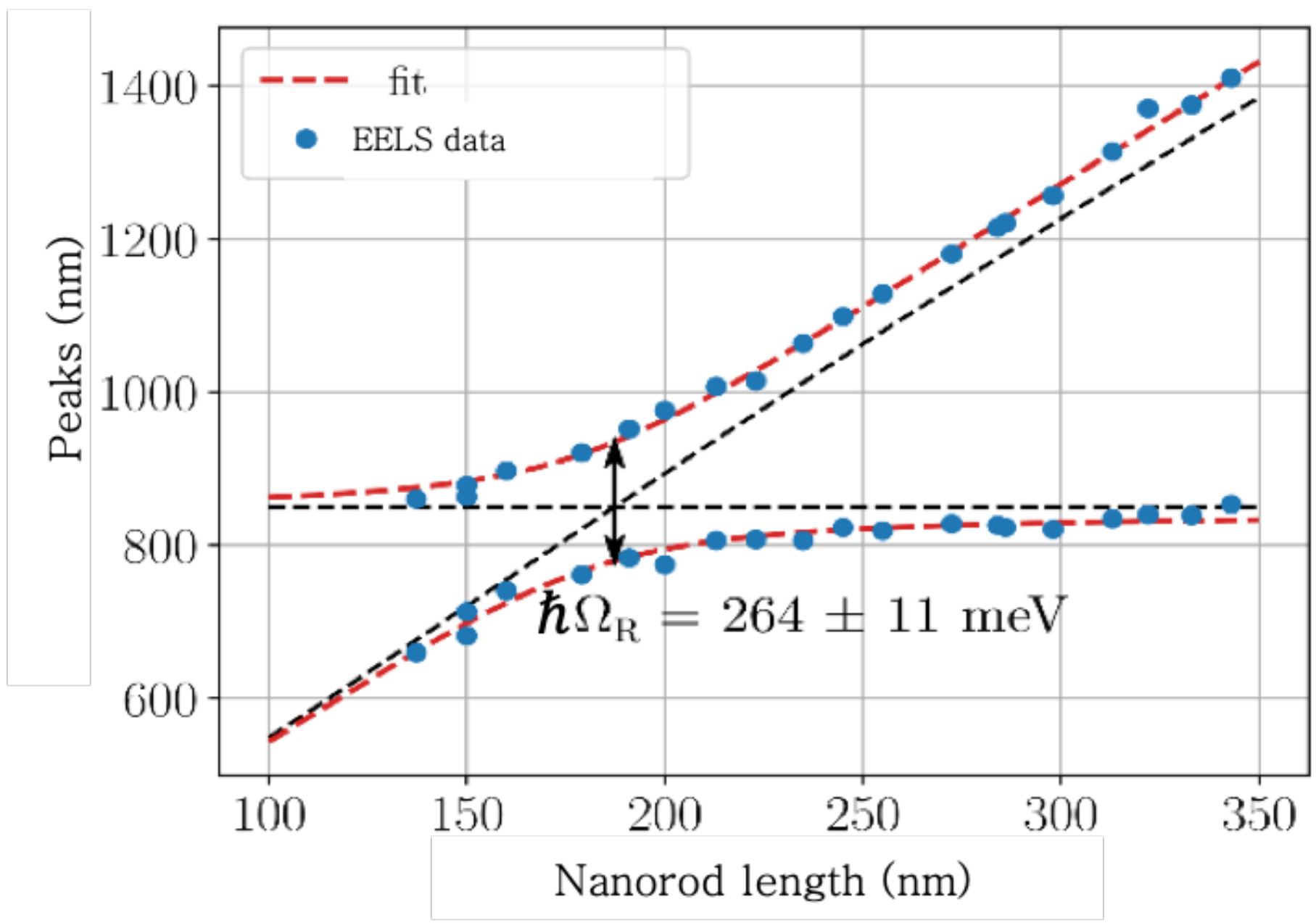

**Figure S8.** Evolution of the EELS peak positions nanorod as a function of the rod length in the cases of dipolar modes.

## REFERENCES

1. Lide, D. R., Ed. CRC handbook of chemistry and physics, 88th ed.; CRC press, 2007.
2. Palik, E. D. (Ed.). (1985). Handbook of optical constants of solids. Academic press.
3. Eizner, E.; Avayu, O.; Ditcovski, R.; Ellenbogen, T. Aluminum nanoantenna complexes for strong coupling between excitons and localized surface plasmons. Nano Lett. 2015, 15, 6215–6221.
4. Lecarme, O.; Sun, Q.; Ueno, K.; Misawa, H. Robust and versatile light absorption at near-infrared wavelengths by plasmonic aluminum nanorods. ACS Photon. 2014, 1, 538–546.
5. Martin, J.; Kociak, M.; Mahfoud, Z.; Proust, J.; Gérard, D.; Plain, J. High-resolution imaging and spectroscopy of multipolar plasmonic resonances in aluminum nanoantennas. Nano Lett. 2014, 14, 5517–5523.